\documentclass[%
 aip,
 amsmath,amssymb,
 reprint,%
]{revtex4-1}

\usepackage{graphicx}
\usepackage{epstopdf}
\usepackage{dcolumn}
\usepackage{bm}

\usepackage[utf8]{inputenc}
\usepackage[T1]{fontenc}
\usepackage{mathptmx}
\usepackage{etoolbox}
\usepackage{color}

\makeatletter
\def\@email#1#2{%
 \endgroup
 \patchcmd{\titleblock@produce}
  {\frontmatter@RRAPformat}
  {\frontmatter@RRAPformat{\produce@RRAP{*#1\href{mailto:#2}{#2}}}\frontmatter@RRAPformat}
  {}{}
}%
\makeatother
\begin{document}

\preprint{AIP/123-QED}

\title{Observing Quantum Coherent Oscillations in a Three-Level Atoms via Electromagnetically Induced Transparency by Two-Dimensional Spectroscopy}

\author{Jing-Yi-Ran Jin}
\affiliation{School of Physics and Astronomy, Applied Optics Beijing Area Major Laboratory, Beijing Normal University, Beijing 100875, China}
\affiliation{Key Laboratory of Multiscale Spin Physics, Ministry of Education, Beijing Normal University, Beijing 100875, China}

\author{Hao-Yue Zhang}
\affiliation{School of Physics and Astronomy, Applied Optics Beijing Area Major Laboratory, Beijing Normal University, Beijing 100875, China}
\affiliation{Key Laboratory of Multiscale Spin Physics, Ministry of Education, Beijing Normal University, Beijing 100875, China}

\author{Yi-Xuan Yao}
\affiliation{School of Physics and Astronomy, Applied Optics Beijing Area Major Laboratory, Beijing Normal University, Beijing 100875, China}
\affiliation{Key Laboratory of Multiscale Spin Physics, Ministry of Education, Beijing Normal University, Beijing 100875, China}

\author{Rong-Hang Chen}
\affiliation{Beijing Computational Science Research Center, Beijing 100084, China}
\affiliation{School of Physics and Astronomy, Applied Optics Beijing Area Major Laboratory, Beijing Normal University, Beijing 100875, China}
\affiliation{Key Laboratory of Multiscale Spin Physics, Ministry of Education, Beijing Normal University, Beijing 100875, China}

\author{Qing Ai}
\email{aiqing@bnu.edu.cn}
\affiliation{School of Physics and Astronomy, Applied Optics Beijing Area Major Laboratory, Beijing Normal University, Beijing 100875, China}
\affiliation{Key Laboratory of Multiscale Spin Physics, Ministry of Education, Beijing Normal University, Beijing 100875, China}

\date{\today}

\begin{abstract}
Two-dimensional electronic spectroscopy (2DES) has high spectral resolution and is a useful tool for studying atom dynamics. In this paper, we show a smallest unit of electromagnetically induced transparency (EIT) for 2DES, that is, a three-level system. It is found that the original main peak is split into four small ones due to the introduction of the EIT. It suggests that the homogeneous broadening of 2DES can be effectively reduced by the EIT. Moreover, in sharp contrast to a constant height, the height of the peaks will manifest a damped oscillation with respect to the population time. It seems that the quantum-beat phenomenon appears. These findings may help us obtain more information about the dynamics of excited states.

\end{abstract}
\maketitle

\section{Introduction}
\label{sec:introduction}
Electromagnetically induced transparency (EIT) is an atomic coherence effect produced by a multi-level atomic system in the presence of an external light field \cite{Scully1997}. It can be specifically described as when the control and probe fields are incident on the system with the two-photon resonance condition, the nonlinear effects of the medium are enhanced and thus the absorption and dispersion properties are significantly altered due to the destructive interference effect between the excitation paths. Because of its unique optical properties, EIT has a wide range of applications in cold atoms \cite{hopkins1997OE,hau1999Nature,shiau2011PRL}, solid materials \cite{schmidt1996OC,zhao1997PRL,kuznetsova2002PRA,baldit2010PRB,fan2019PRA}, nitrogen-vacancy color centers in diamond \cite{wei1999PRA,Hemmer2001OL,acosta2013PRL,Wang2018PRA}, quantum information processing \cite{fleischhauer2000PRL,liu2001Nature,fleischhauer2002PRA,honda2008PRL,hsiao2018PRL,chen2013PRL,kukharchyk2020OE,Wang2023AdP}, metamaterials \cite{jin2013JO,bagci2019PLA,jin2011OC,tassin2009PRL,xu2010OE}, and non-reciprocal optics \cite{Huang2022AdP}, since it was first experimentally observed by Harris' group in 1991 \cite{boller1991PRL}.

Two-dimensional electronic spectroscopy (2DES) is a third-order nonlinear spectral measurement technique \cite{Mukamel1999}, which can reflect the absorption characteristics. Three ultrafast optical coherent pulses interact with matter sequentially, and the collected optical signals containing the evolution information are transformed by double Fourier transforms to obtain 2DES. Compared with widely-used narrowband-excitation and wide-spectrum detection methods, 2DES can have high spectral resolution and time resolution in the case of wideband excitation. Thus, it can capture the quantum coherent phenomenon in the early stage of optical excitation \cite{schlau2011CP,fuller2015ARPC,Zhu2020jpca}. It can measure the homogeneous and inhomogeneous broadening of the sample \cite{sun2023AQT}, study intermolecular interactions \cite{Ito2014jcp}, and also identify the coupling between different transitions by cross peaks \cite{liu2020JPCL,Nagata2007jcp}.
Therefore, it plays a vital role in the investigation of the quantum dot \cite{seiler2018NL,brosseau2020JCP}, two-dimensional materials \cite{moody2015NC,guo2019NP}, perovskite \cite{jha2017AP,monahan2017JPCL,seiler2019NC}, and energy-transfer mechanism of photosynthesis \cite{calhoun2009JPCB,fuller2014NC,Zhu2024nc,romero2014NP,Policht2022sa,ma2017JPCL,thyrhaug2018NC,Silori2023jpcl,Zhu2022jcp,Zhang2021jcp}.
In this paper, we combine EIT with 2DES. Different from the previous work \cite{Deng2024pra}, we apply it on the smallest unit and find that the spectral resolution can be improved. Moreover, considering the change of the excited state population with time when the temperature is constant, it is found that the quantum beat phenomenon occurs, which is more helpful to explore the excited state dynamics.


This paper is organized as follows. In Sec.~\ref{sec:model}, we derive the response functions of rephasing and non-rephasing signals under the established model by using Feynman diagrams. In Sec.~\ref{sec:results}, we show 2DES of rephasing, non-rephasing and absorptive signals in the presence and absence of a control field. We compare 2DES before and after introducing EIT and analyze the underlying physical mechanism of this phenomenon. In Sec.~\ref{sec:conclusion}, we summarize our main findings.

\section{Model}
\label{sec:model}

\begin{figure}
\includegraphics[bb=0 0 550 620, width=9cm]{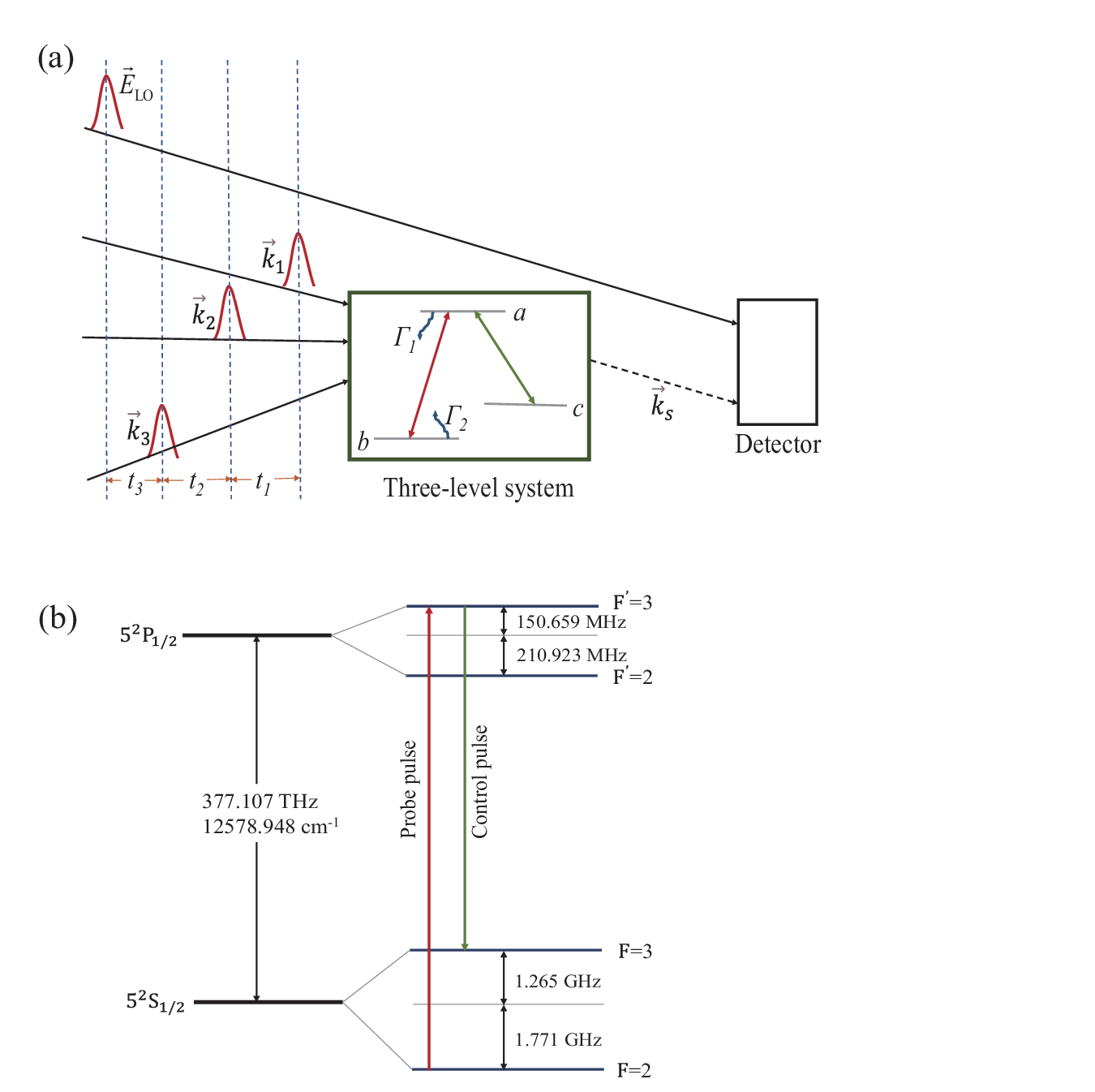}%
\caption{(a) Schematic diagram for 2DES of a three-level atom, where energy levels $b$ and $c$ are respectively the ground and metastable states, and energy level $a$ is the excited state. $\Gamma_{1}$ and $\Gamma_{2}$ represent
the relaxation rates between the energy levels $a$ and $b$. (b) The relevant energy level of ${\rm^{85}Rb}$ atom in D$_1$ transition and the transitions driven by the pulses \cite{Rb85}.}
\label{fig:scheme}
\end{figure}

For the three-level atom as shown in the Fig.~\ref{fig:scheme}, four short pulses in the time domain
compose a detection pulse sequence, and the time intervals between
them are respectively $t_{1}$, $t_{2}$ and $t_{3}$. The first three detection pulses interact with the sample and resonantly probe the transition between the ground state $b$ and the excited state $a$. Third-order nonlinear polarization reads
\begin{eqnarray}
P^{(3)}(t)&=&\int_0^\infty dt_3\int_0^\infty dt_2\int_0^\infty dt_1E(t-t_3)E(t-t_3-t_2)\nonumber\\
&&\times E(t-t_3-t_2-t_1)S^{(3)}(t_3,t_2,t_1),
\end{eqnarray}
where the third-order nonlinear response function is
\begin{eqnarray}
S^{(3)}(t_3,t_2,t_1)&=&(-\frac{i}{\hbar})^3\langle\mu(t_3+t_2+t_1)\times \nonumber\\
&&[\mu(t_2+t_1)[\mu(t_1)[\mu(0),\rho(-\infty)]]]\rangle,
\end{eqnarray}
with $\rho$ the density matrix of the system, $\mu$ the dipole operator. On account of the time ordering, the rotation-wave approximation and the phase-matching condition, we can significantly simplify the computation. And the generated third-order polarization signal will be emitted in the direction along the wave vector $\vec{k}_{s}$, which will be specified below. Since the signal is weak, a pulse $\vec{E}_\textrm{LO}$ in the same direction as $\vec{k}_{s}$ is used as the local oscillating light for heterodyne measurement. In addition, a narrow-band control light is applied in resonance with the transition
between the excited state $a$ and the metastable state $c$. In this way, we construct a simple $\Lambda$-type electromagnetically induced transparent model for the 2DES.

\begin{figure}
\includegraphics[bb=40 20 730 370, width=8.5cm]{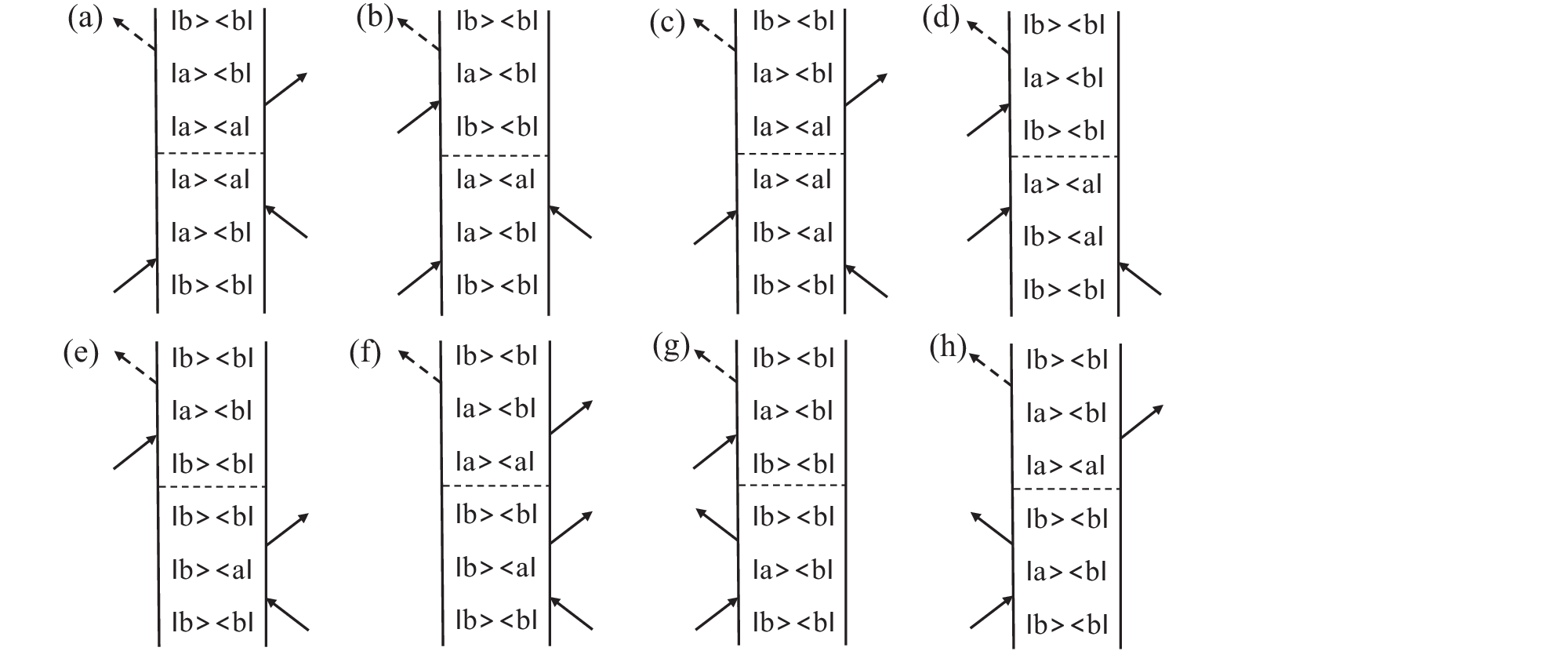}
\caption{Feynman diagrams for the three-level system, with $\tilde{R_1}$ for (a) and (b), $\tilde{R_2}$ for (c) and (d), $\tilde{R_3}$ for (e) and (f), and $\tilde{R_4}$ for (g) and (h). The initial and final states of the population time are separated by a dashed line.}\label{fig:FeynmanDiagram}
\end{figure}

In experiments, there are three types of observable signals, i.e., the absorptive signal
\begin{eqnarray}
S^{(3)}_\textrm{ab}(\omega_{3},t_{2},\omega_{1})
&=&\textrm{Re}\sum_{i=1}^4\tilde{R_i}(\omega_{3},t_{2},\omega_{1}),
\end{eqnarray}
the rephasing signal
\begin{eqnarray}
S^{(3)}_\textrm{rp}(\omega_{3},t_{2},\omega_{1})
&\!=\!&\textrm{Re}[\tilde{R_2}(\omega_{3},t_{2},\omega_{1})+\tilde{R_3}(\omega_{3},t_{2},\omega_{1})],
\end{eqnarray}
the non-rephasing signal
\begin{eqnarray}
S^{(3)}_\textrm{nr}(\omega_{3},t_{2},\omega_{1})
&\!=\!&\textrm{Re}[\tilde{R_1}(\omega_{3},t_{2},\omega_{1})+\tilde{R_4}(\omega_{3},t_{2},\omega_{1})],
\end{eqnarray}
where the different Liouville paths $\tilde{R_{\alpha}}$'s ($\alpha=1,2,3,4$) are schematically illustrated in Fig.~\ref{fig:FeynmanDiagram}.
In the double-sided Feynman diagrams \cite{Mukamel1999}, the arrows represent the action of the probe and signal light fields. The right-headed arrow represents the electric-field component $\mathrm{e}^{-i\omega t+ikr}$, while the left-headed arrow represents the electric-field component $\mathrm{e}^{i\omega t-ikr}$. The arrows pointing towards the system represent the excitation processes, while the arrows pointing away from the system correspond to the deexcitation processes. The last action is derived from the tracing operation, which implies an emitted signal. Since it originates from a physical mechanism different from the previous three, it is denoted by a dashed arrow. Eventually the density matrix will be in a population state.

During the population time $t_{2}$, $\tilde{R_1}$ and $\tilde{R_2}$ represent the initial evolution of the electron in the excited states, while $\tilde{R_3}$ and $\tilde{R_4}$ represent the initial evolution of the electron in the ground state. On the other hand, $\tilde{R_2}$ and $\tilde{R_3}$ indicate that the signal responds
in the direction with the wave vector  $\vec{k}_{s}=-\vec{k}_{1}+\vec{k}_{2}+\vec{k}_{3}$,
while $\tilde{R_1}$ and $\tilde{R_4}$ indicate that the signal responds
in the direction with the wave vector  $\vec{k}_{s}=\vec{k}_{1}-\vec{k}_{2}+\vec{k}_{3}$.

For the $\Lambda$-type three-level atom as considered in Fig.~\ref{fig:scheme}, the Hamiltonian can be expressed as
\begin{eqnarray}
H&=&H_{0}+H_\textrm{int},\\
H_{0}&=&\sum_{j=a,b,c}\hbar\omega_{j}\vert j\rangle\langle j\vert,\\
H_\textrm{int}&=&-\frac{\hbar}{2}\Omega e^{-i\nu_{c}t}\vert a\rangle\langle c\vert+\textrm{h.c.},
\end{eqnarray}
where $\hbar\omega_{j}$ is the energy of the state $\vert j\rangle$ with $\hbar$ being the reduced Planck constant, $\Omega=\mu_{ac}\varepsilon_{c}/\hbar$ is the Rabi frequency, $\nu_{c}$ and $\varepsilon_{c}$ are respectively the frequency and amplitude of the control field,
$\mu_{ac}=\langle a\vert\mu \vert c\rangle$ is the transition dipole moment. Because the probe field is so strong and short that the control field and the longitudinal and transverse relaxations can be neglected when the probe fields are present. However, in the time evolution between the two sequential probe pulses, the effect of the probe field can be ignored, and both the effects of the control field and the relaxations need to be taken into account.

In the presence of control light, the reduced density matrix of the system is governed by the quantum master equation \cite{Breuer2002}
\begin{eqnarray}
\dot{\rho}&=&-\frac{i}{\hbar}\left[H,\rho\right]-\Gamma_1\mathcal{L}(\vert b\rangle\langle a\vert)\rho-\Gamma_2\mathcal{L}(\vert a\rangle\langle b\vert)\rho\nonumber\\
&&-\sum_{\alpha=a,b,c}\gamma_\alpha^{(0)}\mathcal{L}(\vert \alpha\rangle\langle \alpha\vert)\rho,
\end{eqnarray}
where the Lindblad operator
\begin{equation}
\mathcal{L}(A)\rho=\frac{1}{2}\{A^\dagger A,\rho\}- A\rho A^\dagger,
\end{equation}
$\{A,\rho\}=A\rho+\rho A$ is the anti-commutator. $\Gamma_{1}$ and $\Gamma_{2}$ are respectively the downhill and uphill relaxation rates between the energy levels $a$ and $b$. $\gamma_\alpha^{(0)}$ ($\alpha=a,b,c$) is the pure-dephasing noise of the state $\vert \alpha\rangle$. By considering natural radiation damping and phase relaxation, it is possible to more accurately describe the behavior of the system in the excited state \cite{Tanimura2006jpsj,Tanimura1989jpsj,Takagahara1977jpsj}. It is worth noting that since we consider the coupling between the system and the environment to be weak,  the perturbation theory can be applied. Because the correlation time of the environmental fluctuations is very short, the Markovian and rotating-wave approximation is valid. However, anomalies associated with the ohmic spin-boson system exist even at finite temperatures \cite{Leggett1987rmp}. And since in some chemical and biological systems, the environment is complex and strongly coupled to the system, and the correlation time of the environment is comparable to the time scale of the system dynamics, the hierarchical equations of motion (HEOM) theory reproduces more-reliable results for the quantum dynamics and thus the spectra \cite{Tanimura2014jcp,Tanimura2020jcp,Koyanagi2024jcp,Tanimura2015jcp,Koyanagi2024jcp2}.

%

By transforming to the interaction picture, after some algebra,
we can obtain a set of differential equations for the matrix elements of $\rho$ as
\begin{eqnarray}
\dot{\rho}_{bb}&=&\Gamma_{1}\rho_{aa}-\Gamma_{2}\rho_{bb}
,\\
\dot{\rho}_{ba}&=&-\frac{i}{2}\Omega\rho_{bc}-\gamma_{1}\rho_{ba}
,\\
\dot{\rho}_{bc}&=&-\frac{i}{2}\Omega\rho_{ba}-\gamma_{2}\rho_{bc}
,\\
\dot{\rho}_{ab}&=&\frac{i}{2}\Omega\rho_{cb}-\gamma_{1}\rho_{ab}
,\\
\dot{\rho}_{aa}&=&\frac{i}{2}\Omega(\rho_{ca}-\rho_{ac})-\Gamma_{1}\rho_{aa}+\Gamma_{2}\rho_{bb}
,\\
\dot{\rho}_{ac}&=&\frac{i}{2}\Omega(\rho_{cc}-\rho_{aa})-\gamma_{3}\rho_{ac}
,\\
\dot{\rho}_{cb}&=&\frac{i}{2}\Omega\rho_{ab}-\gamma_{2}\rho_{cb}
,\\
\dot{\rho}_{ca}&=&\frac{i}{2}\Omega(\rho_{aa}-\rho_{cc})-\gamma_{3}\rho_{ca}
,\\
\dot{\rho}_{cc}&=&\frac{i}{2}\Omega(\rho_{ac}-\rho_{ca}),
\end{eqnarray}
where for simplicity we introduce
\begin{eqnarray}
\gamma_{1} & =&\frac{1}{2}(\Gamma_{1}+\gamma_{a}^{(0)}+\Gamma_{2}+\gamma_{b}^{(0)}),\\
\gamma_{2} & =&\frac{1}{2}(\Gamma_{2}+\gamma_{b}^{(0)}+\gamma_{c}^{(0)}),\\
\gamma_{3} & =&\frac{1}{2}(\Gamma_{1}+\gamma_{a}^{(0)}+\gamma_{c}^{(0)}).
\end{eqnarray}

\begin{figure}
\includegraphics[bb=10 10 390 300,width=8.5cm]{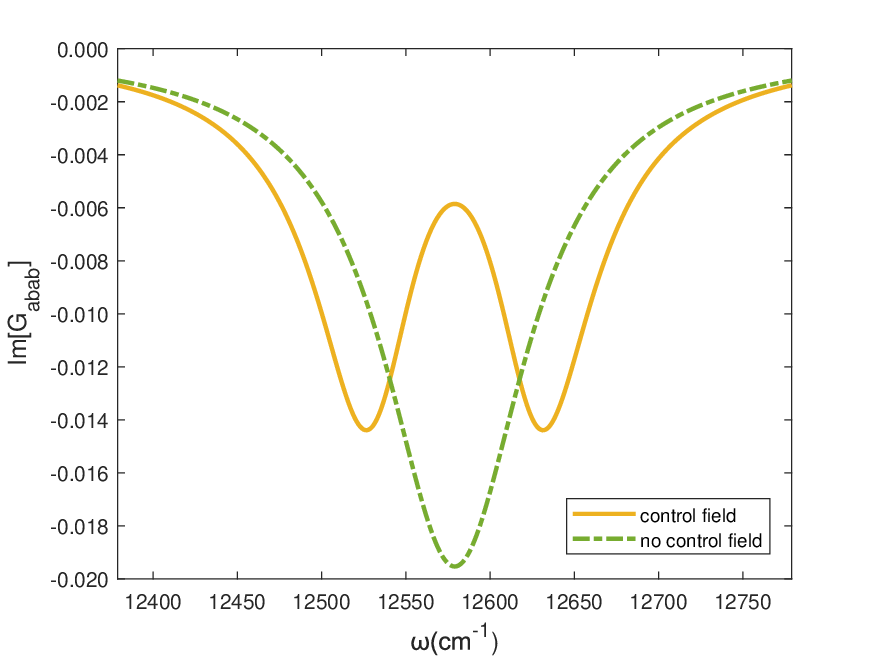}
\caption{Spectral plots of the imaginary part of $\mathcal{G}_{ab,ab}(\omega)$ in the presence or absence of the control field. The parameters are $\omega_{ab}=12578.948$~cm$^{-1}$, $\gamma_1=\gamma_3=2\pi\times0.2446~\text{THz}$ \cite{Yan2022pra}, $\Gamma_1=2\pi\times5.75~\text{GHz}$, $\gamma_2=2\pi\times0.1~\text{THz}$ and $\Omega=3~\text{THz}$.}
\label{fig:G_abab}
\end{figure}

The relevant Green's functions are explicitly obtained as
\begin{eqnarray}\label{eq:GF}
\mathrm{\mathcal{G}}_{ba,ba}(\omega_{1})&= & \frac{4(\omega_{1}-i\gamma_{2}-\omega_{ab})}{4(\omega_{1}-i\gamma_{1}-\omega_{ab})(\omega_{1}-i\gamma_{2}-\omega_{ab})-\Omega^{2}},\nonumber\\
\mathrm{\mathcal{G}}_{ab,ab}(\omega_{3})&= & \frac{4(\omega_{3}+i\gamma_{2}-\omega_{ab})}{4(\omega_{3}+i\gamma_{1}-\omega_{ab})(\omega_{3}+i\gamma_{2}-\omega_{ab})-\Omega^{2}},\nonumber\\
\end{eqnarray}
where the Green's functions $\mathcal{G}_{ba,ba}(\omega_{1})$ and
$\mathcal{G}_{ab,ab}(\omega_{3})$ have double peaks and a higher minimum with the presence of the control field than without, as shown in Fig.~\ref{fig:G_abab}. By using the Green's functions, the rephasing and non-rephasing signals can be explicitly given as
\begin{eqnarray}
S^{(3)}_\textrm{rp}(\omega_{3},t_{2},\omega_{1})=&&\textrm{Re}[\mathcal{G}_{ab,ab}(\omega_{3})\mathcal{G}_{aa,aa}(t_{2})\mathcal{G}_{ba,ba}(\omega_{1})\nonumber\\
&&+\mathcal{G}_{ab,ab}(\omega_{3})\mathcal{G}_{bb,aa}(t_{2})\mathcal{G}_{ba,ba}(\omega_{1})\nonumber\\
&&+\mathcal{G}_{ab,ab}(\omega_{3})\mathcal{G}_{aa,bb}(t_{2})\mathcal{G}_{ba,ba}(\omega_{1})\nonumber\\
&&+\mathcal{G}_{ab,ab}(\omega_{3})\mathcal{G}_{bb,bb}(t_{2})\mathcal{G}_{ba,ba}(\omega_{1})],\label{eq:rp}\\
S^{(3)}_\textrm{nr}(\omega_{3},t_{2},\omega_{1})=&&\textrm{Re}[\mathcal{G}_{ab,ab}(\omega_{3})\mathcal{G}_{aa,aa}(t_{2})\mathcal{G}_{ab,ab}(\omega_{1})\nonumber\\
&&+\mathcal{G}_{ab,ab}(\omega_{3})\mathcal{G}_{bb,aa}(t_{2})\mathcal{G}_{ab,ab}(\omega_{1})\nonumber\\
&&+\mathcal{G}_{ab,ab}(\omega_{3})\mathcal{G}_{aa,bb}(t_{2})\mathcal{G}_{ab,ab}(\omega_{1})\nonumber\\
&&+\mathcal{G}_{ab,ab}(\omega_{3})\mathcal{G}_{bb,bb}(t_{2})\mathcal{G}_{ab,ab}(\omega_{1})].\label{eq:nr}
\end{eqnarray}

\begin{figure}
\includegraphics[bb=0 0 410 630,width=8.5cm]{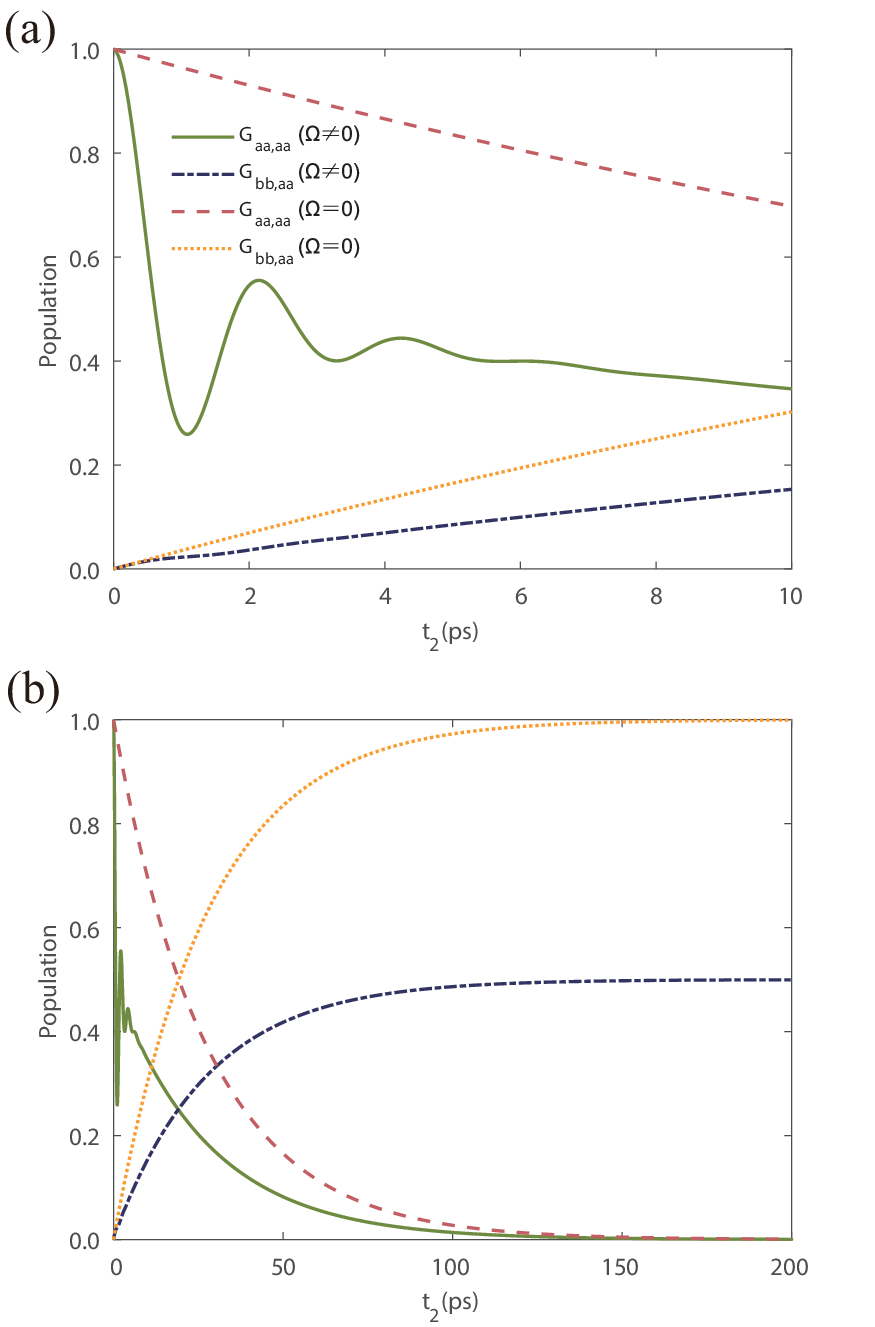}
\caption{In the presence and absence of a control field, population dynamics of $|a\rangle$ and $|b\rangle$. Sub-figure (a) is an enlarged view of (b) when $t_2$ is shortened. The other parameters are the same as in Fig.~\ref{fig:G_abab}.}
\label{fig:population}
\end{figure}

We consider a case when the system is prepared at $|a\rangle$ after the first two pulses, cf. $\tilde{R_1}$ and $\tilde{R_2}$ in Fig.~\ref{fig:FeynmanDiagram}. After $t_2$, the probability of the system at $|a\rangle$ and $|b\rangle$ are
\begin{eqnarray}
\mathcal{G}_{aa,aa}(t_2)\simeq && e^{-\frac{\gamma_{3}}{2}t_2}\left[\frac{1}{2}\cos(\widetilde{\Omega}t_2)\frac{A_{2}\Gamma_{2}+\Omega^{2}}{A_{1}}\right.\nonumber\\
&&\left.+\frac{1}{4}\sin(\widetilde{\Omega}t_2)\frac{A_{2}\Gamma_{2}\gamma_{3}+(\gamma_{3}-2\Gamma_{1})\Omega^{2}}{\widetilde{\Omega}A_{1}}\right]\label{eq:G_aaaa}\\
&&+\frac{\Gamma_{2}}{2(\Gamma_{1}+\Gamma_{2})}+\frac{\Gamma_{1}(2A_{1}-\Omega^{2})e^{-(\Gamma_{1}+\Gamma_{2})t_2}}{2(\Gamma_{1}+\Gamma_{2})A_{1}},\nonumber\\
\mathcal{G}_{bb,aa}(t_2)\simeq && e^{-\frac{\gamma_{3}}{2}t_2}\left[\frac{1}{2}\cos(\widetilde{\Omega}t_2)\frac{A_2\Gamma_{1}}{A_{1}}\right.\nonumber\\
&&\left.+\frac{1}{4}\sin(\widetilde{\Omega}t_2)\frac{\Gamma_{1}(A_2\gamma_{3}+2\Omega^{2})}{\widetilde{\Omega}A_{1}}\right]\label{eq:G_bbaa}\\
&&+\frac{\Gamma_{1}}{2(\Gamma_{1}+\Gamma_{2})}+\frac{\Gamma_{1}(\Omega^{2}-2A_{1})e^{-(\Gamma_{1}+\Gamma_{2})t_2}}{2(\Gamma_{1}+\Gamma_{2})A_{1}},\nonumber
\end{eqnarray}
where
\begin{eqnarray}  A_{1}&=&(\Gamma_{1}+\Gamma_{2})(\Gamma_{1}+\Gamma_{2}-\gamma_{3})+\Omega^{2},\\
  A_{2}&=&\Gamma_{1}+\Gamma_{2}-\gamma_{3},\\
  \widetilde{\Omega}&=&\frac{1}{2}\sqrt{4\Omega^{2}-\gamma_{3}^{2}}.
\end{eqnarray}
When $\Omega=0$, Eqs.~\eqref{eq:G_aaaa} and \eqref{eq:G_bbaa} can be simplified as
\begin{eqnarray}
\mathcal{G}_{aa,aa}(t_2)=&& \frac{1}{\Gamma_1+\Gamma_2}\left[\Gamma_2+\Gamma_1e^{-(\Gamma_1+\Gamma_2)t_2}\right],\label{eq:G_aaaa_noCF}\\
\mathcal{G}_{bb,aa}(t_2)=&&
\frac{\Gamma_1}{\Gamma_1+\Gamma_2}\left[1-e^{-(\Gamma_1+\Gamma_2)t_2}\right]\label{eq:G_bbaa_noCF}.
\end{eqnarray}
In Fig.~\ref{fig:population}, we numerically plot $\mathcal{G}_{aa,aa}(t_2)$ and $\mathcal{G}_{bb,aa}(t_2)$ in the presence of the control field. It can be found that after a time scale with $2/\gamma_3$, the coherent oscillations vanish, $\mathcal{G}_{aa,aa}(t_2)$ and $\mathcal{G}_{bb,aa}(t_2)$ finally approach a steady state $\Gamma_2/2(\Gamma_1+\Gamma_2)$ and $\Gamma_1/2(\Gamma_1+\Gamma_2)$. Since the total population is conserved, half of the population is transferred to $\vert c\rangle$. This results in the nearly-identical population transfer from $\vert a\rangle$ to $\vert b\rangle$ and from $\vert a\rangle$ to $\vert c\rangle$ because $\Gamma_1\gg\Gamma_2$. However, when the control field disappears, the coherent oscillation no longer exists, and the population in $|a\rangle$ and $|b\rangle$ change monotonically until they reach a steady state of $\Gamma_2/(\Gamma_1+\Gamma_2)$ and $\Gamma_1/(\Gamma_1+\Gamma_2)$.

Similarly, when the system is prepared at $|b\rangle$ after the first two pulses, cf. $\tilde{R_3}$ and $\tilde{R_4}$ in Fig.~\ref{fig:FeynmanDiagram}, the probability of the system at $|a\rangle$ and $|b\rangle$ are
\begin{eqnarray}\label{eq:rho_bb}
\mathcal{G}_{aa,bb}(t_2)\simeq\frac{\Gamma_{2}-e^{-(\Gamma_{1}+\Gamma_{2})t_2}\Gamma_{2}}{\Gamma_{1}+\Gamma_{2}},\label{eq:G_aabb}\\
\mathcal{G}_{bb,bb}(t_2)\simeq\frac{\Gamma_{1}+e^{-(\Gamma_{1}+\Gamma_{2})t_2}\Gamma_{2}}{\Gamma_{1}+\Gamma_{2}}\label{eq:G_bbbb}.
\end{eqnarray}
Since there is no control field applied to the transition between $|a\rangle$ and $|b\rangle$, there is no coherent oscillation of $\mathcal{G}_{aa,bb}(t_2)$ and $\mathcal{G}_{bb,bb}(t_2)$ as compared to $\mathcal{G}_{aa,aa}(t_2)$ and $\mathcal{G}_{bb,aa}(t_2)$. However, after a time scale with $1/(\Gamma_1+\Gamma_2)$, they finally approach a steady state $\Gamma_2/(\Gamma_1+\Gamma_2)$ and $\Gamma_1/(\Gamma_1+\Gamma_2)$ as implied by the detailed balance.

\section{Results}
\label{sec:results}

\begin{figure}
\includegraphics[bb=10 10 370 315, width=8.5cm]{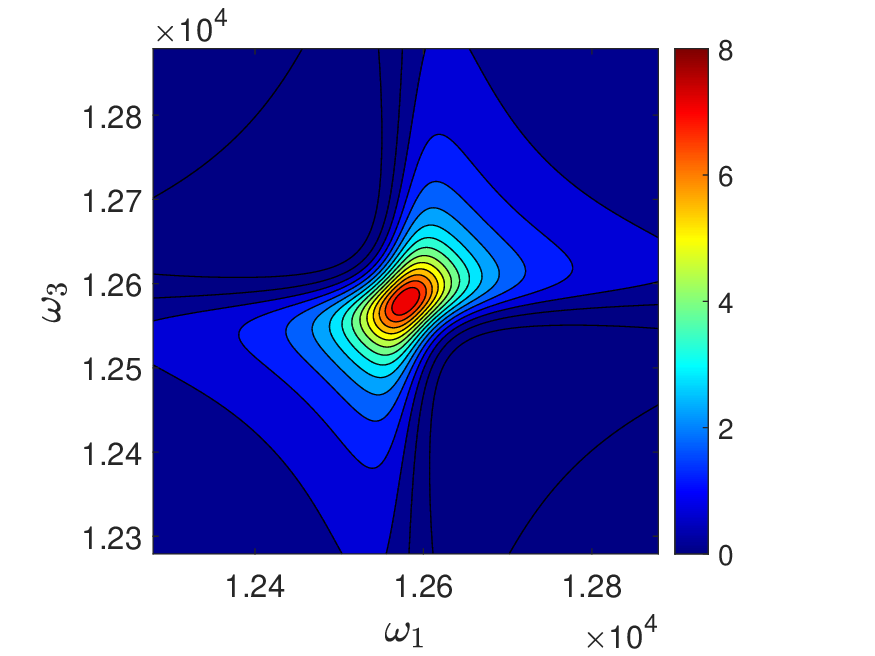}
\caption{In the absence of the control field, the 2D spectra of the rephasing signal. The parameters are the same as in Fig.~\ref{fig:G_abab}.}
\label{fig:rephasing_no}
\end{figure}

\begin{figure}
\includegraphics[bb=20 20 400 370, width=8.5cm]{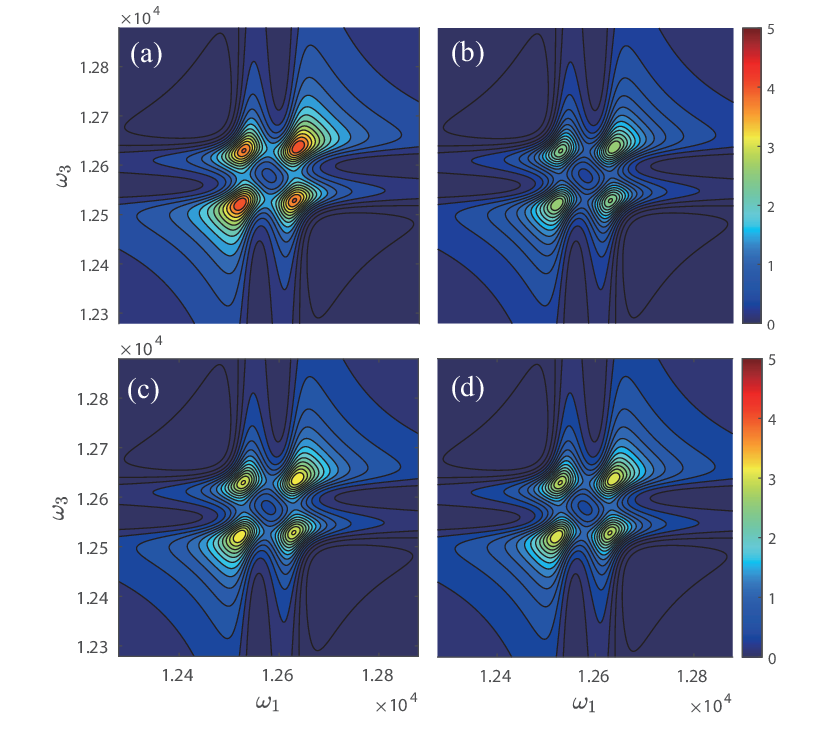}
\caption{In the presence of the control field, the 2D spectra of the rephasing signal for (a) $t_{2}=0$, (b) $t_{2}=1$~ps, (c) $t_{2}=2$~ps, and (d) $t_{2}=150$~ps. The parameters are the same as in Fig.~\ref{fig:G_abab}.}
\label{fig:rephasing_control}
\end{figure}

We consider that both the control and the probe pulses are linearly polarized and their polarization directions are perpendicular to each other \cite{Zhang2008ol}. The linewidth of the probe pulse is assumed to be of the
 order of THz \cite{Seiler2017jcp,Wang2022pra}. Rb atoms are in an environment of 160~$^{\circ}\text{C}$ \cite{Yan2022pra}, according to the detailed-balance conditions, $\Gamma_2/\Gamma_1\approx 10^{-18}$. The probe pulse can induce the D$_1$ atomic transition $5^{2}S_{1/2},F=2 \rightarrow 5^{2}P_{1/2},F'=3$, while the control pulse can induce the D$_1$ atomic transition $5^{2}S_{1/2},F=3 \rightarrow 5^{2}P_{1/2},F'=3$ \cite{Geng2014njp}. First of all, we plot the rephasing signal without the control field in Fig.~\ref{fig:rephasing_no}. As shown, there is only one peak at $\omega_{ab}=\omega_{a}-\omega_{b}$. As the population time $t_2$ elapses, both the position and the height of the peak remains the same according to Eqs.~\eqref{eq:G_aaaa_noCF}, \eqref{eq:G_bbaa_noCF}, \eqref{eq:G_aabb}, \eqref{eq:G_bbbb} and \eqref{eq:rp}. This is because the number of atoms in the energy levels $a$ and $b$ is conserved. In other words, $\mathcal{G}_{aa,aa}(t_2)+\mathcal{G}_{bb,aa}(t_2)=1$ and $\mathcal{G}_{aa,bb}(t_2)+\mathcal{G}_{bb,bb}(t_2)=1$. When $t_2\gg(\Gamma_1+\Gamma_2)^{-1}$, the population transfer effectively ends. However, when the control field is applied, as shown in Fig.~\ref{fig:rephasing_control}, due to the EIT effect, the original one large peak is split into four small ones with widths significantly narrower than the original one, and thus the homogeneous broadening of the spectrum is reduced. As shown in Fig.~\ref{fig:G_abab}, the imaginary part of the Green functions in $t_3$ have been changed from single trough at $\omega_3=\omega_{ab}$ to double troughs at
\begin{eqnarray}
&&\omega_3=\omega_{ab}\nonumber\\
&&\pm\frac{1}{2}\sqrt{\frac{\Omega(\gamma_1+\gamma_2)\sqrt{4\gamma_1\gamma_2+\Omega^{2}}-\gamma_2(4\gamma_1\gamma_2+\Omega^{2})}{\gamma_1}}, \end{eqnarray}
according to Eq.~\eqref{eq:GF}. The imaginary part of Green's function in $t_1$ will likewise change from one trough to two troughs.
The reason for this phenomenon is given as follows. The larger the magnitude of the negative imaginary Green function is, the more the probe field is absorbed by the medium and thus the bigger the transition probability is. In the absence of the control field, the particle will only transit between the states $|b\rangle$ and $|a\rangle$. When the control field is applied, there are many different contributions to the transition paths, which start from $|b\rangle$ and end at $|a\rangle$, such as $|b\rangle\rightarrow|a\rangle\rightarrow|c\rangle\rightarrow|a\rangle$, $|b\rangle\rightarrow|a\rangle\rightarrow|c\rangle\rightarrow|a\rangle\rightarrow|c\rangle\rightarrow|a\rangle$ and so on. There will be interference between these transition paths, and the absorption of the probe field is the sum of the transition probability amplitudes of all paths. When the signs of the transition probability amplitudes are opposite to each other and thus cancel with each other, EIT is formed. Although both the shapes and the positions of these peaks will not be changed as $t_2$ is enlarged, the heights of the peaks decline and rise with $t_2\sim2/\gamma_3$, which is the quantum-beat phenomenon, until steady state according to Eqs.~\eqref{eq:G_aaaa}, \eqref{eq:G_bbaa}, \eqref{eq:G_aabb}, \eqref{eq:G_bbbb} and \eqref{eq:rp}.

\begin{figure}
\includegraphics[bb=10 10 370 315, width=8.5cm]{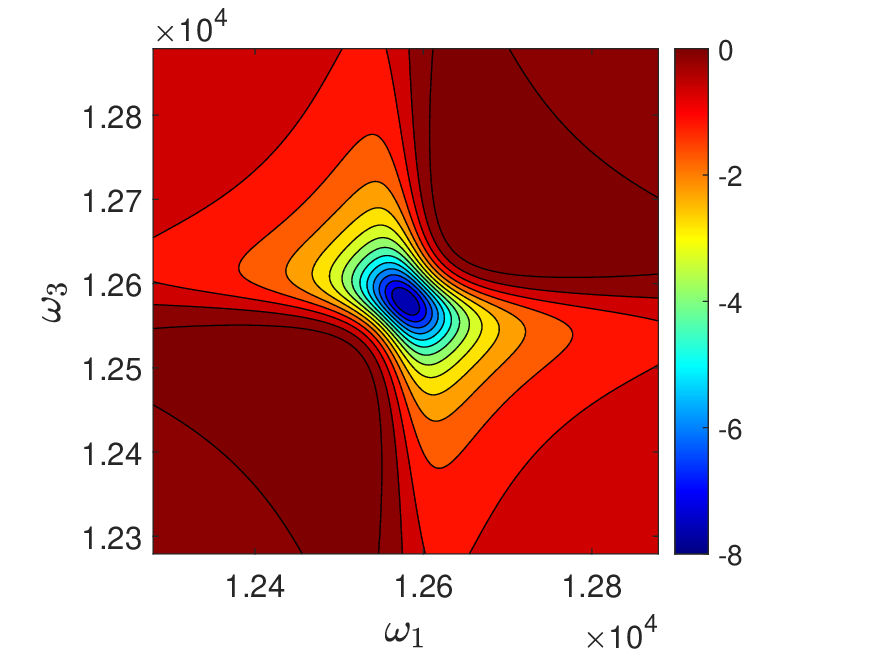}
\caption{In the absence of the control field, the 2D spectra of the non-rephasing signal. The other parameters are the same as in Fig.~\ref{fig:G_abab}.}
\label{fig:nonrephasing_no}
\end{figure}

\begin{figure}
\includegraphics[bb=20 20 390 350, width=8.5cm]{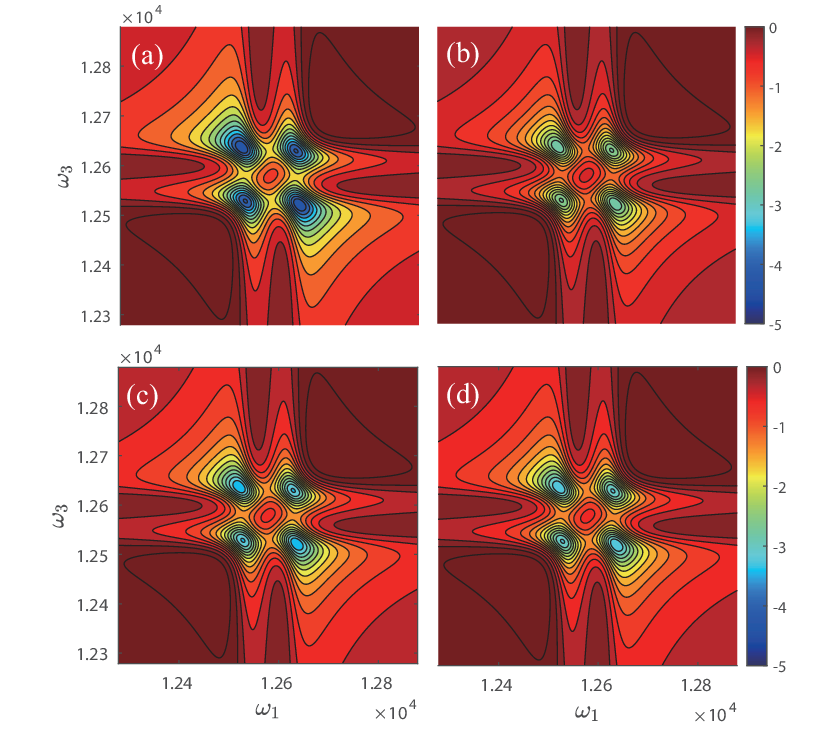}
\caption{In the presence of the control field, the 2D spectra of the non-rephasing signal for (a) $t_{2}=0$, (b) $t_{2}=1$~ps, (c) $t_{2}=2$~ps, and (d) $t_{2}=150$~ps. The other parameters are the same as in Fig.~\ref{fig:G_abab}.}
\label{fig:nonrephasing_control}
\end{figure}

Then, we plot the non-rephasing signal in the absence of the control field in Fig.~\ref{fig:nonrephasing_no}. Because the states in the durations of $t_1$ and $t_3$ are the same for $\tilde{R_1}$ and $\tilde{R_4}$, a trough appears at $\omega_{1}=\omega_{3}=\omega_{ab}$ in the two-dimensional spectroscopy in this case. The depth of the trough will not change with the extension of $t_2$ time, as can be seen from Eqs.~\eqref{eq:G_aaaa_noCF}, \eqref{eq:G_bbaa_noCF}, \eqref{eq:G_aabb}, \eqref{eq:G_bbbb} and \eqref{eq:nr}. When the control field is added, that is, EIT is introduced, one trough in the original two-dimensional spectroscopy will be split into four small troughs, as shown in Fig.~\ref{fig:nonrephasing_control}. Equations~\eqref{eq:G_aaaa}, \eqref{eq:G_bbaa}, \eqref{eq:G_aabb}, \eqref{eq:G_bbbb} and \eqref{eq:nr} imply that as $t_2$ time increases, the population in the $a$ and $b$ states will carry on the coherent oscillations with decreasing amplitude. After about $150$~ps, the populations of both $a$ and $b$ states become stable. Hence, both the position and shape of the troughs do not change in the spectroscopy, but their depths oscillate with the passage of $t_2$ time.

\begin{figure}
\includegraphics[bb=15 10 360 315, width=8.5cm]{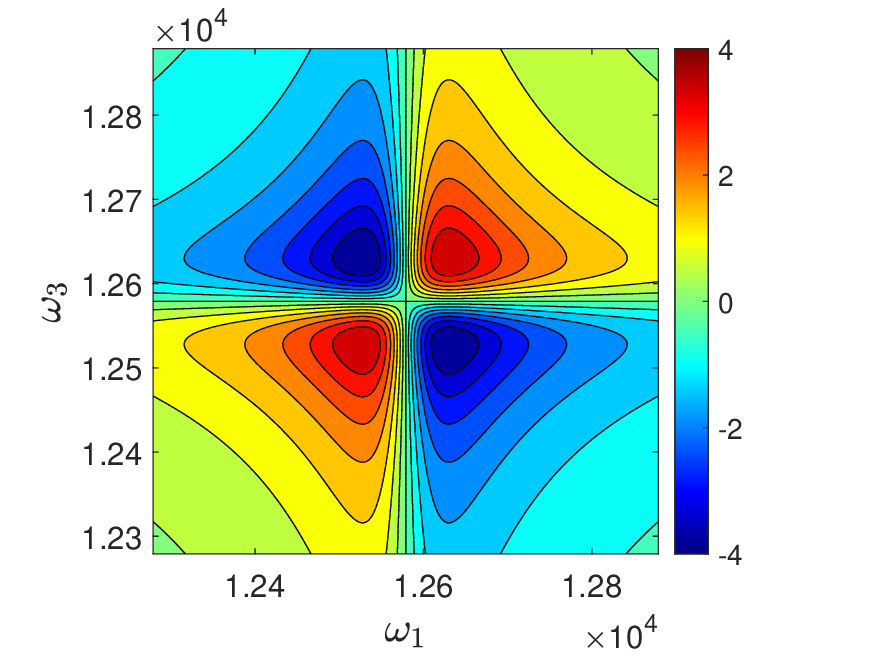}
\caption{In the absence of the control field, the 2D spectra of the absorptive signal. The other parameters are the same as in Fig.~\ref{fig:G_abab}.}
\label{fig:absorptive_no}
\end{figure}

\begin{figure}
\includegraphics[bb=20 10 390 350, width=8.5cm]{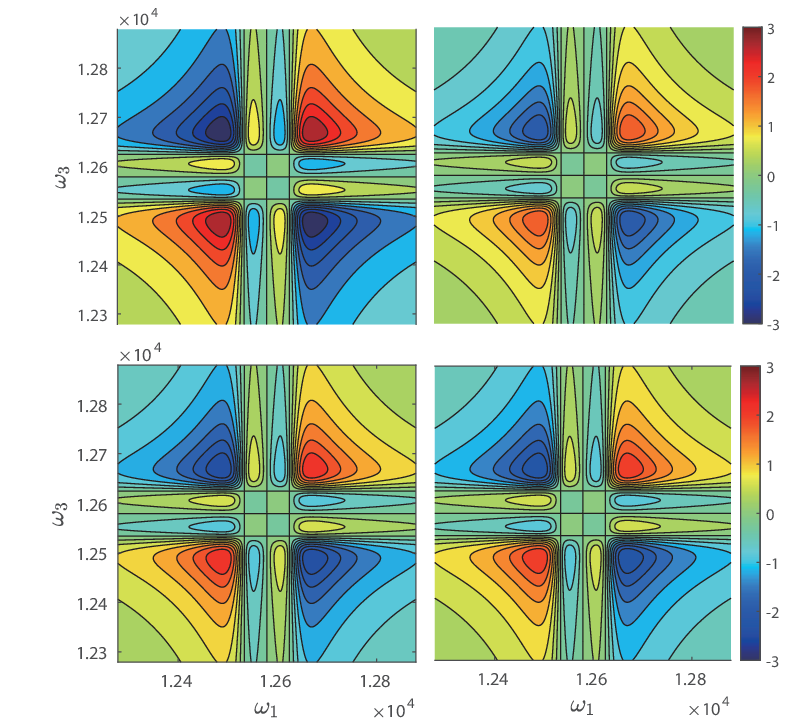}
\caption{In the presence of the control field, the 2D spectra of the absorptive signal for (a) $t_{2}=0$, (b) $t_{2}=1$~ps, (c) $t_{2}=2$~ps, and (d) $t_{2}=150$~ps. The other parameters are the same as in Fig.~\ref{fig:G_abab}.}
\label{fig:absorptive_control}
\end{figure}

Summing the rephasing and non-rephasing signals, the total absorptive 2D spectroscopy is obtained, which is more complicated as can be seen from Fig.~\ref{fig:absorptive_no}. There are two peaks along the diagonal line, and two troughs along the anti-diagonal line. Moreover, with the increase of $t_{2}$ time, the atoms in the $a$ and $b$ energy levels transition to each other, but the overall height of the peaks and the depth of the troughs remain the same. When the control field is applied, as shown in Fig.~\ref{fig:absorptive_control}, the homogeneous broadening of all peaks and troughs are effectively reduced. In addition to the above two sets of main peaks and troughs, two new sets of peaks and troughs emerge, i.e., eight narrow peaks and troughs between one main peak and trough, and four small ones at the center. Similar to the rephasing and non-rephasing signals, quantum-beat phenomenon also appears along with the time evolution, and both the peaks and troughs show damped oscillations, eventually arriving at the steady state in the long-time limit. Neither the number nor the positions of the peaks and troughs in the 2DES change due to the elapse of $t_2$. It implies that both the rephasing and non-rephasing spectroscopy possess the shape over $\omega_1$ and $\omega_3$, although the magnitude changes over time.

\section{Conclusion}
\label{sec:conclusion}

In this paper, we apply EIT which can induce constructive and destructive interference to two-dimensional spectroscopy. It is found that in the rephasing and non-rephasing signals, the original one large peak (trough) is split into four small peaks (troughs), which indicates that the homogeneous broadening in the spectroscopy will be reduced. With the evolution of the population time, the height of the peaks (the depth of the troughs) will rise and fall. In other words, the quantum-beat phenomena occur, until a steady value in the long-time limit due to the transition of the atoms. But both the position and shape of the peak (trough) will not change during the evolution. The time evolution of the absorptive signal will show a similar characteristic although its spectroscopy is much more complicated. This means that EIT can improve the resolution of 2DES, and this method is promising to obtain more information about excited state dynamics.

\textit{Acknowledgments.}---
This work is supported by Innovation Program for Quantum Science and Technology under Grant No.~2023ZD0300200, the National Natural Science Foundation of China under Grant No.~62461160263, Beijing Natural Science Foundation under Grant No.~1202017, and Beijing Normal University under Grant No.~2022129.


\begin{thebibliography}{48}%
\makeatletter
\providecommand \@ifxundefined [1]{%
 \@ifx{#1\undefined}
}%
\providecommand \@ifnum [1]{%
 \ifnum #1\expandafter \@firstoftwo
 \else \expandafter \@secondoftwo
 \fi
}%
\providecommand \@ifx [1]{%
 \ifx #1\expandafter \@firstoftwo
 \else \expandafter \@secondoftwo
 \fi
}%
\providecommand \natexlab [1]{#1}%
\providecommand \enquote  [1]{``#1''}%
\providecommand \bibnamefont  [1]{#1}%
\providecommand \bibfnamefont [1]{#1}%
\providecommand \citenamefont [1]{#1}%
\providecommand \href@noop [0]{\@secondoftwo}%
\providecommand \href [0]{\begingroup \@sanitize@url \@href}%
\providecommand \@href[1]{\@@startlink{#1}\@@href}%
\providecommand \@@href[1]{\endgroup#1\@@endlink}%
\providecommand \@sanitize@url [0]{\catcode `\\12\catcode `\$12\catcode
  `\&12\catcode `\#12\catcode `\^12\catcode `\_12\catcode `\%12\relax}%
\providecommand \@@startlink[1]{}%
\providecommand \@@endlink[0]{}%
\providecommand \url  [0]{\begingroup\@sanitize@url \@url }%
\providecommand \@url [1]{\endgroup\@href {#1}{\urlprefix }}%
\providecommand \urlprefix  [0]{URL }%
\providecommand \Eprint [0]{\href }%
\providecommand \doibase [0]{http://dx.doi.org/}%
\providecommand \selectlanguage [0]{\@gobble}%
\providecommand \bibinfo  [0]{\@secondoftwo}%
\providecommand \bibfield  [0]{\@secondoftwo}%
\providecommand \translation [1]{[#1]}%
\providecommand \BibitemOpen [0]{}%
\providecommand \bibitemStop [0]{}%
\providecommand \bibitemNoStop [0]{.\EOS\space}%
\providecommand \EOS [0]{\spacefactor3000\relax}%
\providecommand \BibitemShut  [1]{\csname bibitem#1\endcsname}%
\let\auto@bib@innerbib\@empty
\bibitem [{\citenamefont {Scully}\ and\ \citenamefont
  {Zubairy}(1997)}]{Scully1997}%
  \BibitemOpen
  \bibfield  {author} {\bibinfo {author} {\bibfnamefont {M.~O.}\ \bibnamefont
  {Scully}}\ and\ \bibinfo {author} {\bibfnamefont {M.~S.}\ \bibnamefont
  {Zubairy}},\ }\href@noop {} {\emph {\bibinfo {title} {Quantum Optics}}}\
  (\bibinfo  {publisher} {Cambridge University Press, New York},\ \bibinfo
  {year} {1997})\BibitemShut {NoStop}%
\bibitem [{\citenamefont {Hopkins}\ \emph {et~al.}(1997)\citenamefont
  {Hopkins}, \citenamefont {Usadi}, \citenamefont {Chen},\ and\ \citenamefont
  {Durrant}}]{hopkins1997OE}%
  \BibitemOpen
  \bibfield  {author} {\bibinfo {author} {\bibfnamefont {S.~A.}\ \bibnamefont
  {Hopkins}}, \bibinfo {author} {\bibfnamefont {E.}~\bibnamefont {Usadi}},
  \bibinfo {author} {\bibfnamefont {H.~X.}\ \bibnamefont {Chen}}, \ and\
  \bibinfo {author} {\bibfnamefont {A.~V.}\ \bibnamefont {Durrant}},\
  }\bibfield  {title} {\enquote {\bibinfo {title} {{Electromagnetically induced
  transparency of laser-cooled rubidium atoms in three-level $\Lambda$-type
  systems}},}\ }\href {\doibase 10.1016/S0030-4018(97)00030-8} {\bibfield
  {journal} {\bibinfo  {journal} {Opt. Commun.}\ }\textbf {\bibinfo {volume}
  {138}},\ \bibinfo {pages} {185} (\bibinfo {year} {1997})}\BibitemShut
  {NoStop}%
\bibitem [{\citenamefont {Hau}\ \emph {et~al.}(1999)\citenamefont {Hau},
  \citenamefont {Harris}, \citenamefont {Dutton},\ and\ \citenamefont
  {Behroozi}}]{hau1999Nature}%
  \BibitemOpen
  \bibfield  {author} {\bibinfo {author} {\bibfnamefont {L.~V.}\ \bibnamefont
  {Hau}}, \bibinfo {author} {\bibfnamefont {S.~E.}\ \bibnamefont {Harris}},
  \bibinfo {author} {\bibfnamefont {Z.}~\bibnamefont {Dutton}}, \ and\ \bibinfo
  {author} {\bibfnamefont {C.~H.}\ \bibnamefont {Behroozi}},\ }\bibfield
  {title} {\enquote {\bibinfo {title} {Light speed reduction to 17 metres per
  second in an ultracold atomic gas},}\ }\href {\doibase 10.1038/17561}
  {\bibfield  {journal} {\bibinfo  {journal} {Nature}\ }\textbf {\bibinfo
  {volume} {397}},\ \bibinfo {pages} {594} (\bibinfo {year}
  {1999})}\BibitemShut {NoStop}%
\bibitem [{\citenamefont {Shiau}\ \emph {et~al.}(2011)\citenamefont {Shiau},
  \citenamefont {Wu}, \citenamefont {Lin},\ and\ \citenamefont
  {Chen}}]{shiau2011PRL}%
  \BibitemOpen
  \bibfield  {author} {\bibinfo {author} {\bibfnamefont {B.-W.}\ \bibnamefont
  {Shiau}}, \bibinfo {author} {\bibfnamefont {M.-C.}\ \bibnamefont {Wu}},
  \bibinfo {author} {\bibfnamefont {C.-C.}\ \bibnamefont {Lin}}, \ and\
  \bibinfo {author} {\bibfnamefont {Y.-C.}\ \bibnamefont {Chen}},\ }\bibfield
  {title} {\enquote {\bibinfo {title} {Low-light-level cross-phase modulation
  with double slow light pulses},}\ }\href {\doibase
  10.1103/PhysRevLett.106.193006} {\bibfield  {journal} {\bibinfo  {journal}
  {Phys. Rev. Lett.}\ }\textbf {\bibinfo {volume} {106}},\ \bibinfo {pages}
  {193006} (\bibinfo {year} {2011})}\BibitemShut {NoStop}%
\bibitem [{\citenamefont {Schmidt}\ and\ \citenamefont
  {Imamo?lu}(1996)}]{schmidt1996OC}%
  \BibitemOpen
  \bibfield  {author} {\bibinfo {author} {\bibfnamefont {H.}~\bibnamefont
  {Schmidt}}\ and\ \bibinfo {author} {\bibfnamefont {A.}~\bibnamefont
  {Imamo\v{g}lu}},\ }\bibfield  {title} {\enquote {\bibinfo {title} {Nonlinear
  optical devices based on a transparency in semiconductor intersubband
  transitions},}\ }\href {\doibase 10.1016/0030-4018(96)00354-9} {\bibfield
  {journal} {\bibinfo  {journal} {Opt. Commun.}\ }\textbf {\bibinfo {volume}
  {131}},\ \bibinfo {pages} {333} (\bibinfo {year} {1996})}\BibitemShut
  {NoStop}%
\bibitem [{\citenamefont {Zhao}\ \emph {et~al.}(1997)\citenamefont {Zhao},
  \citenamefont {Wu}, \citenamefont {Ham}, \citenamefont {Kim},\ and\
  \citenamefont {Awad}}]{zhao1997PRL}%
  \BibitemOpen
  \bibfield  {author} {\bibinfo {author} {\bibfnamefont {Y.}~\bibnamefont
  {Zhao}}, \bibinfo {author} {\bibfnamefont {C.-K.}\ \bibnamefont {Wu}},
  \bibinfo {author} {\bibfnamefont {B.-S.}\ \bibnamefont {Ham}}, \bibinfo
  {author} {\bibfnamefont {M.~K.}\ \bibnamefont {Kim}}, \ and\ \bibinfo
  {author} {\bibfnamefont {E.}~\bibnamefont {Awad}},\ }\bibfield  {title}
  {\enquote {\bibinfo {title} {Microwave induced transparency in ruby},}\
  }\href {\doibase 10.1103/PhysRevLett.79.641} {\bibfield  {journal} {\bibinfo
  {journal} {Phys. Rev. Lett.}\ }\textbf {\bibinfo {volume} {79}},\ \bibinfo
  {pages} {641} (\bibinfo {year} {1997})}\BibitemShut {NoStop}%
\bibitem [{\citenamefont {Kuznetsova}\ \emph {et~al.}(2002)\citenamefont
  {Kuznetsova}, \citenamefont {Kocharovskaya}, \citenamefont {Hemmer},\ and\
  \citenamefont {Scully}}]{kuznetsova2002PRA}%
  \BibitemOpen
  \bibfield  {author} {\bibinfo {author} {\bibfnamefont {E.}~\bibnamefont
  {Kuznetsova}}, \bibinfo {author} {\bibfnamefont {O.}~\bibnamefont
  {Kocharovskaya}}, \bibinfo {author} {\bibfnamefont {P.}~\bibnamefont
  {Hemmer}}, \ and\ \bibinfo {author} {\bibfnamefont {M.~O.}\ \bibnamefont
  {Scully}},\ }\bibfield  {title} {\enquote {\bibinfo {title} {Atomic
  interference phenomena in solids with a long-lived spin coherence},}\ }\href
  {\doibase 10.1103/PhysRevA.66.063802} {\bibfield  {journal} {\bibinfo
  {journal} {Phys. Rev. A}\ }\textbf {\bibinfo {volume} {66}},\ \bibinfo
  {pages} {063802} (\bibinfo {year} {2002})}\BibitemShut {NoStop}%
\bibitem [{\citenamefont {Baldit}\ \emph {et~al.}(2010)\citenamefont {Baldit},
  \citenamefont {Bencheikh}, \citenamefont {Monnier}, \citenamefont
  {Briaudeau}, \citenamefont {Levenson}, \citenamefont {Crozatier},
  \citenamefont {Lorger{\'e}}, \citenamefont {Bretenaker}, \citenamefont
  {Le~Gou{\"e}t}, \citenamefont {Guillot-No{\"e}l},\ and\ \citenamefont
  {Goldner}}]{baldit2010PRB}%
  \BibitemOpen
  \bibfield  {author} {\bibinfo {author} {\bibfnamefont {E.}~\bibnamefont
  {Baldit}}, \bibinfo {author} {\bibfnamefont {K.}~\bibnamefont {Bencheikh}},
  \bibinfo {author} {\bibfnamefont {P.}~\bibnamefont {Monnier}}, \bibinfo
  {author} {\bibfnamefont {S.}~\bibnamefont {Briaudeau}}, \bibinfo {author}
  {\bibfnamefont {J.~A.}\ \bibnamefont {Levenson}}, \bibinfo {author}
  {\bibfnamefont {V.}~\bibnamefont {Crozatier}}, \bibinfo {author}
  {\bibfnamefont {I.}~\bibnamefont {Lorger{\'e}}}, \bibinfo {author}
  {\bibfnamefont {F.}~\bibnamefont {Bretenaker}}, \bibinfo {author}
  {\bibfnamefont {J.~L.}\ \bibnamefont {Le~Gou{\"e}t}}, \bibinfo {author}
  {\bibfnamefont {O.}~\bibnamefont {Guillot-No{\"e}l}}, \ and\ \bibinfo
  {author} {\bibfnamefont {Ph.}\ \bibnamefont {Goldner}},\ }\bibfield  {title}
  {\enquote {\bibinfo {title} {{Identification of $\Lambda$-like systems in
  Er$^{3+}$: Y$_2$SiO$_5$ and observation of electromagnetically induced
  transparency}},}\ }\href {\doibase 10.1103/PhysRevB.81.144303} {\bibfield
  {journal} {\bibinfo  {journal} {Phys. Rev. B}\ }\textbf {\bibinfo {volume}
  {81}},\ \bibinfo {pages} {144303} (\bibinfo {year} {2010})}\BibitemShut
  {NoStop}%
\bibitem [{\citenamefont {Fan}\ \emph {et~al.}(2019)\citenamefont {Fan},
  \citenamefont {Kagalwala}, \citenamefont {Polyakov}, \citenamefont
  {Migdall},\ and\ \citenamefont {Goldschmidt}}]{fan2019PRA}%
  \BibitemOpen
  \bibfield  {author} {\bibinfo {author} {\bibfnamefont {H.-Q.}\ \bibnamefont
  {Fan}}, \bibinfo {author} {\bibfnamefont {K.~H.}\ \bibnamefont {Kagalwala}},
  \bibinfo {author} {\bibfnamefont {S.~V.}\ \bibnamefont {Polyakov}}, \bibinfo
  {author} {\bibfnamefont {A.~L.}\ \bibnamefont {Migdall}}, \ and\ \bibinfo
  {author} {\bibfnamefont {E.~A.}\ \bibnamefont {Goldschmidt}},\ }\bibfield
  {title} {\enquote {\bibinfo {title} {Electromagnetically induced transparency
  in inhomogeneously broadened solid media},}\ }\href {\doibase
  10.1103/PhysRevA.99.053821} {\bibfield  {journal} {\bibinfo  {journal} {Phys.
  Rev. A}\ }\textbf {\bibinfo {volume} {99}},\ \bibinfo {pages} {053821}
  (\bibinfo {year} {2019})}\BibitemShut {NoStop}%
\bibitem [{\citenamefont {Wei}\ and\ \citenamefont
  {Manson}(1999)}]{wei1999PRA}%
  \BibitemOpen
  \bibfield  {author} {\bibinfo {author} {\bibfnamefont {C.-J.}\ \bibnamefont
  {Wei}}\ and\ \bibinfo {author} {\bibfnamefont {N.~B.}\ \bibnamefont
  {Manson}},\ }\bibfield  {title} {\enquote {\bibinfo {title} {Observation of
  the dynamic stark effect on electromagnetically induced transparency},}\
  }\href {\doibase 10.1103/PhysRevA.60.2540} {\bibfield  {journal} {\bibinfo
  {journal} {Phys. Rev. A}\ }\textbf {\bibinfo {volume} {60}},\ \bibinfo
  {pages} {2540} (\bibinfo {year} {1999})}\BibitemShut {NoStop}%
\bibitem [{\citenamefont {Hemmer}\ \emph {et~al.}(2001)\citenamefont {Hemmer},
  \citenamefont {Turukhin}, \citenamefont {Shahriar},\ and\ \citenamefont
  {Musser}}]{Hemmer2001OL}%
  \BibitemOpen
  \bibfield  {author} {\bibinfo {author} {\bibfnamefont {P.~R.}\ \bibnamefont
  {Hemmer}}, \bibinfo {author} {\bibfnamefont {A.~V.}\ \bibnamefont
  {Turukhin}}, \bibinfo {author} {\bibfnamefont {M.~S.}\ \bibnamefont
  {Shahriar}}, \ and\ \bibinfo {author} {\bibfnamefont {J.~A.}\ \bibnamefont
  {Musser}},\ }\bibfield  {title} {\enquote {\bibinfo {title} {Raman-excited
  spin coherences in nitrogen-vacancy color centers in diamond},}\ }\href
  {\doibase 10.1364/OL.26.000361} {\bibfield  {journal} {\bibinfo  {journal}
  {Opt. Lett.}\ }\textbf {\bibinfo {volume} {26}},\ \bibinfo {pages} {361}
  (\bibinfo {year} {2001})}\BibitemShut {NoStop}%
\bibitem [{\citenamefont {Acosta}\ \emph {et~al.}(2013)\citenamefont {Acosta},
  \citenamefont {Jensen}, \citenamefont {Santori}, \citenamefont {Budker},\
  and\ \citenamefont {Beausoleil}}]{acosta2013PRL}%
  \BibitemOpen
  \bibfield  {author} {\bibinfo {author} {\bibfnamefont {V.~M.}\ \bibnamefont
  {Acosta}}, \bibinfo {author} {\bibfnamefont {K.}~\bibnamefont {Jensen}},
  \bibinfo {author} {\bibfnamefont {C.}~\bibnamefont {Santori}}, \bibinfo
  {author} {\bibfnamefont {D.}~\bibnamefont {Budker}}, \ and\ \bibinfo {author}
  {\bibfnamefont {R.~G.}\ \bibnamefont {Beausoleil}},\ }\bibfield  {title}
  {\enquote {\bibinfo {title} {Electromagnetically induced transparency in a
  diamond spin ensemble enables all-optical electromagnetic field sensing},}\
  }\href {\doibase 10.1103/PhysRevLett.110.213605} {\bibfield  {journal}
  {\bibinfo  {journal} {Phys. Rev. Lett.}\ }\textbf {\bibinfo {volume} {110}},\
  \bibinfo {pages} {213605} (\bibinfo {year} {2013})}\BibitemShut {NoStop}%
\bibitem [{\citenamefont {Wang}\ \emph {et~al.}(2018)\citenamefont {Wang},
  \citenamefont {Qiu}, \citenamefont {Chu}, \citenamefont {Zhang},
  \citenamefont {Cai}, \citenamefont {Ai},\ and\ \citenamefont
  {Deng}}]{Wang2018PRA}%
  \BibitemOpen
  \bibfield  {author} {\bibinfo {author} {\bibfnamefont {Y.-Y.}\ \bibnamefont
  {Wang}}, \bibinfo {author} {\bibfnamefont {J.}~\bibnamefont {Qiu}}, \bibinfo
  {author} {\bibfnamefont {Y.-Q.}\ \bibnamefont {Chu}}, \bibinfo {author}
  {\bibfnamefont {M.}~\bibnamefont {Zhang}}, \bibinfo {author} {\bibfnamefont
  {J.-M.}\ \bibnamefont {Cai}}, \bibinfo {author} {\bibfnamefont
  {Q.}~\bibnamefont {Ai}}, \ and\ \bibinfo {author} {\bibfnamefont {F.-G.}\
  \bibnamefont {Deng}},\ }\bibfield  {title} {\enquote {\bibinfo {title} {Dark
  state polarizing a nuclear spin in the vicinity of a nitrogen-vacancy
  center},}\ }\href {\doibase 10.1103/PhysRevA.97.042313} {\bibfield  {journal}
  {\bibinfo  {journal} {Phys. Rev. A}\ }\textbf {\bibinfo {volume} {97}},\
  \bibinfo {pages} {042313} (\bibinfo {year} {2018})}\BibitemShut {NoStop}%
\bibitem [{\citenamefont {Fleischhauer}\ and\ \citenamefont
  {Lukin}(2000)}]{fleischhauer2000PRL}%
  \BibitemOpen
  \bibfield  {author} {\bibinfo {author} {\bibfnamefont {M.}~\bibnamefont
  {Fleischhauer}}\ and\ \bibinfo {author} {\bibfnamefont {M.~D.}\ \bibnamefont
  {Lukin}},\ }\bibfield  {title} {\enquote {\bibinfo {title} {Dark-state
  polaritons in electromagnetically induced transparency},}\ }\href {\doibase
  10.1103/PhysRevLett.84.5094} {\bibfield  {journal} {\bibinfo  {journal}
  {Phys. Rev. Lett.}\ }\textbf {\bibinfo {volume} {84}},\ \bibinfo {pages}
  {5094} (\bibinfo {year} {2000})}\BibitemShut {NoStop}%
\bibitem [{\citenamefont {Liu}\ \emph {et~al.}(2001)\citenamefont {Liu},
  \citenamefont {Dutton}, \citenamefont {Behroozi},\ and\ \citenamefont
  {Hau}}]{liu2001Nature}%
  \BibitemOpen
  \bibfield  {author} {\bibinfo {author} {\bibfnamefont {C.}~\bibnamefont
  {Liu}}, \bibinfo {author} {\bibfnamefont {Z.}~\bibnamefont {Dutton}},
  \bibinfo {author} {\bibfnamefont {C.~H.}\ \bibnamefont {Behroozi}}, \ and\
  \bibinfo {author} {\bibfnamefont {L.~V.}\ \bibnamefont {Hau}},\ }\bibfield
  {title} {\enquote {\bibinfo {title} {Observation of coherent optical
  information storage in an atomic medium using halted light pulses},}\ }\href
  {\doibase 10.1038/35054017} {\bibfield  {journal} {\bibinfo  {journal}
  {Nature}\ }\textbf {\bibinfo {volume} {409}},\ \bibinfo {pages} {490}
  (\bibinfo {year} {2001})}\BibitemShut {NoStop}%
\bibitem [{\citenamefont {Fleischhauer}\ and\ \citenamefont
  {Lukin}(2002)}]{fleischhauer2002PRA}%
  \BibitemOpen
  \bibfield  {author} {\bibinfo {author} {\bibfnamefont {M.}~\bibnamefont
  {Fleischhauer}}\ and\ \bibinfo {author} {\bibfnamefont {M.~D.}\ \bibnamefont
  {Lukin}},\ }\bibfield  {title} {\enquote {\bibinfo {title} {Quantum memory
  for photons: Dark-state polaritons},}\ }\href {\doibase
  10.1103/PhysRevA.65.022314} {\bibfield  {journal} {\bibinfo  {journal} {Phys.
  Rev. A}\ }\textbf {\bibinfo {volume} {65}},\ \bibinfo {pages} {022314}
  (\bibinfo {year} {2002})}\BibitemShut {NoStop}%
\bibitem [{\citenamefont {Honda}\ \emph {et~al.}(2008)\citenamefont {Honda},
  \citenamefont {Akamatsu}, \citenamefont {Arikawa}, \citenamefont {Yokoi},
  \citenamefont {Akiba}, \citenamefont {Nagatsuka}, \citenamefont {Tanimura},
  \citenamefont {Furusawa},\ and\ \citenamefont {Kozuma}}]{honda2008PRL}%
  \BibitemOpen
  \bibfield  {author} {\bibinfo {author} {\bibfnamefont {K.}~\bibnamefont
  {Honda}}, \bibinfo {author} {\bibfnamefont {D.}~\bibnamefont {Akamatsu}},
  \bibinfo {author} {\bibfnamefont {M.}~\bibnamefont {Arikawa}}, \bibinfo
  {author} {\bibfnamefont {Y.}~\bibnamefont {Yokoi}}, \bibinfo {author}
  {\bibfnamefont {K.}~\bibnamefont {Akiba}}, \bibinfo {author} {\bibfnamefont
  {S.}~\bibnamefont {Nagatsuka}}, \bibinfo {author} {\bibfnamefont
  {T.}~\bibnamefont {Tanimura}}, \bibinfo {author} {\bibfnamefont
  {A.}~\bibnamefont {Furusawa}}, \ and\ \bibinfo {author} {\bibfnamefont
  {M.}~\bibnamefont {Kozuma}},\ }\bibfield  {title} {\enquote {\bibinfo {title}
  {Storage and retrieval of a squeezed vacuum},}\ }\href {\doibase
  10.1103/PhysRevLett.100.093601} {\bibfield  {journal} {\bibinfo  {journal}
  {Phys. Rev. Lett.}\ }\textbf {\bibinfo {volume} {100}},\ \bibinfo {pages}
  {093601} (\bibinfo {year} {2008})}\BibitemShut {NoStop}%
\bibitem [{\citenamefont {Hsiao}\ \emph {et~al.}(2018)\citenamefont {Hsiao},
  \citenamefont {Tsai}, \citenamefont {Chen}, \citenamefont {Lin},
  \citenamefont {Hung}, \citenamefont {Lee}, \citenamefont {Chen},
  \citenamefont {Chen}, \citenamefont {Yu},\ and\ \citenamefont
  {Chen}}]{hsiao2018PRL}%
  \BibitemOpen
  \bibfield  {author} {\bibinfo {author} {\bibfnamefont {Y.-F.}\ \bibnamefont
  {Hsiao}}, \bibinfo {author} {\bibfnamefont {P.-J.}\ \bibnamefont {Tsai}},
  \bibinfo {author} {\bibfnamefont {H.-S.}\ \bibnamefont {Chen}}, \bibinfo
  {author} {\bibfnamefont {S.-X.}\ \bibnamefont {Lin}}, \bibinfo {author}
  {\bibfnamefont {C.-C.}\ \bibnamefont {Hung}}, \bibinfo {author}
  {\bibfnamefont {C.-H.}\ \bibnamefont {Lee}}, \bibinfo {author} {\bibfnamefont
  {Y.-H.}\ \bibnamefont {Chen}}, \bibinfo {author} {\bibfnamefont {Y.-F.}\
  \bibnamefont {Chen}}, \bibinfo {author} {\bibfnamefont {I.~A.}\ \bibnamefont
  {Yu}}, \ and\ \bibinfo {author} {\bibfnamefont {Y.-C.}\ \bibnamefont
  {Chen}},\ }\bibfield  {title} {\enquote {\bibinfo {title} {Highly efficient
  coherent optical memory based on electromagnetically induced transparency},}\
  }\href {\doibase 10.1103/PhysRevLett.120.183602} {\bibfield  {journal}
  {\bibinfo  {journal} {Phys. Rev. Lett.}\ }\textbf {\bibinfo {volume} {120}},\
  \bibinfo {pages} {183602} (\bibinfo {year} {2018})}\BibitemShut {NoStop}%
\bibitem [{\citenamefont {Chen}\ \emph {et~al.}(2013)\citenamefont {Chen},
  \citenamefont {Lee}, \citenamefont {Wang}, \citenamefont {Du}, \citenamefont
  {Chen}, \citenamefont {Chen},\ and\ \citenamefont {Yu}}]{chen2013PRL}%
  \BibitemOpen
  \bibfield  {author} {\bibinfo {author} {\bibfnamefont {Y.-H.}\ \bibnamefont
  {Chen}}, \bibinfo {author} {\bibfnamefont {M.-J.}\ \bibnamefont {Lee}},
  \bibinfo {author} {\bibfnamefont {I.-Chung}\ \bibnamefont {Wang}}, \bibinfo
  {author} {\bibfnamefont {S.}~\bibnamefont {Du}}, \bibinfo {author}
  {\bibfnamefont {Y.-F.}\ \bibnamefont {Chen}}, \bibinfo {author}
  {\bibfnamefont {Y.-C.}\ \bibnamefont {Chen}}, \ and\ \bibinfo {author}
  {\bibfnamefont {I.~A.}\ \bibnamefont {Yu}},\ }\bibfield  {title} {\enquote
  {\bibinfo {title} {Coherent optical memory with high storage efficiency and
  large fractional delay},}\ }\href {\doibase 10.1103/PhysRevLett.110.083601}
  {\bibfield  {journal} {\bibinfo  {journal} {Phys. Rev. Lett.}\ }\textbf
  {\bibinfo {volume} {110}},\ \bibinfo {pages} {083601} (\bibinfo {year}
  {2013})}\BibitemShut {NoStop}%
\bibitem [{\citenamefont {Kukharchyk}\ \emph {et~al.}(2020)\citenamefont
  {Kukharchyk}, \citenamefont {Sholokhov}, \citenamefont {Morozov},
  \citenamefont {Korableva}, \citenamefont {Kalachev},\ and\ \citenamefont
  {Bushev}}]{kukharchyk2020OE}%
  \BibitemOpen
  \bibfield  {author} {\bibinfo {author} {\bibfnamefont {N.}~\bibnamefont
  {Kukharchyk}}, \bibinfo {author} {\bibfnamefont {D.}~\bibnamefont
  {Sholokhov}}, \bibinfo {author} {\bibfnamefont {O.}~\bibnamefont {Morozov}},
  \bibinfo {author} {\bibfnamefont {S.~L.}\ \bibnamefont {Korableva}}, \bibinfo
  {author} {\bibfnamefont {A.~A.}\ \bibnamefont {Kalachev}}, \ and\ \bibinfo
  {author} {\bibfnamefont {P.~A.}\ \bibnamefont {Bushev}},\ }\bibfield  {title}
  {\enquote {\bibinfo {title} {{Electromagnetically induced transparency in a
  mono-isotopic $^{167}$Er: $^{7}$LiYF$_{4}$ crystal below 1 Kelvin: microwave
  photonics approach}},}\ }\href {\doibase 10.1364/OE.400222} {\bibfield
  {journal} {\bibinfo  {journal} {Opt. Express}\ }\textbf {\bibinfo {volume}
  {28}},\ \bibinfo {pages} {29166} (\bibinfo {year} {2020})}\BibitemShut
  {NoStop}%
\bibitem [{\citenamefont {Wang}\ \emph {et~al.}(2023)\citenamefont {Wang},
  \citenamefont {He}, \citenamefont {Lu}, \citenamefont {Wang}, \citenamefont
  {Ai},\ and\ \citenamefont {Wang}}]{Wang2023AdP}%
  \BibitemOpen
  \bibfield  {author} {\bibinfo {author} {\bibfnamefont {J.}~\bibnamefont
  {Wang}}, \bibinfo {author} {\bibfnamefont {W.-T.}\ \bibnamefont {He}},
  \bibinfo {author} {\bibfnamefont {C.-W.}\ \bibnamefont {Lu}}, \bibinfo
  {author} {\bibfnamefont {Y.-Y.}\ \bibnamefont {Wang}}, \bibinfo {author}
  {\bibfnamefont {Q.}~\bibnamefont {Ai}}, \ and\ \bibinfo {author}
  {\bibfnamefont {H.-B.}\ \bibnamefont {Wang}},\ }\bibfield  {title} {\enquote
  {\bibinfo {title} {Controlled-not gate based on the rydberg states of surface
  electrons},}\ }\href {\doibase 10.1002/andp.202300138} {\bibfield  {journal}
  {\bibinfo  {journal} {Ann. Phys. (Berlin)}\ }\textbf {\bibinfo {volume}
  {535}},\ \bibinfo {pages} {2300138} (\bibinfo {year} {2023})}\BibitemShut
  {NoStop}%
\bibitem [{\citenamefont {Jin}\ \emph {et~al.}(2013)\citenamefont {Jin},
  \citenamefont {Zhang}, \citenamefont {Zhang}, \citenamefont {Lee},\ and\
  \citenamefont {Rhee}}]{jin2013JO}%
  \BibitemOpen
  \bibfield  {author} {\bibinfo {author} {\bibfnamefont {X.-R.}\ \bibnamefont
  {Jin}}, \bibinfo {author} {\bibfnamefont {Y.-Q.}\ \bibnamefont {Zhang}},
  \bibinfo {author} {\bibfnamefont {S.}~\bibnamefont {Zhang}}, \bibinfo
  {author} {\bibfnamefont {Y.-P.}\ \bibnamefont {Lee}}, \ and\ \bibinfo
  {author} {\bibfnamefont {J.-Y.}\ \bibnamefont {Rhee}},\ }\bibfield  {title}
  {\enquote {\bibinfo {title} {Polarization-independent electromagnetically
  induced transparency-like effects in stacked metamaterials based on
  {Fabry--Perot} resonance},}\ }\href {\doibase 10.1088/2040-8978/15/12/125104}
  {\bibfield  {journal} {\bibinfo  {journal} {J. Optics}\ }\textbf {\bibinfo
  {volume} {15}},\ \bibinfo {pages} {125104} (\bibinfo {year}
  {2013})}\BibitemShut {NoStop}%
\bibitem [{\citenamefont {Bagci}\ and\ \citenamefont
  {Akaoglu}(2019)}]{bagci2019PLA}%
  \BibitemOpen
  \bibfield  {author} {\bibinfo {author} {\bibfnamefont {F.}~\bibnamefont
  {Bagci}}\ and\ \bibinfo {author} {\bibfnamefont {B.}~\bibnamefont
  {Akaoglu}},\ }\bibfield  {title} {\enquote {\bibinfo {title} {Transmission
  control by asymmetric electromagnetically induced transparency-like
  metamaterials in transverse electromagnetic waveguide},}\ }\href {\doibase
  10.1016/j.physleta.2019.126000} {\bibfield  {journal} {\bibinfo  {journal}
  {Phys. Lett. A}\ }\textbf {\bibinfo {volume} {383}},\ \bibinfo {pages}
  {126000} (\bibinfo {year} {2019})}\BibitemShut {NoStop}%
\bibitem [{\citenamefont {Jin}\ \emph {et~al.}(2011)\citenamefont {Jin},
  \citenamefont {Lu}, \citenamefont {Zheng}, \citenamefont {Lee}, \citenamefont
  {Rhee}, \citenamefont {Kim},\ and\ \citenamefont {Jang}}]{jin2011OC}%
  \BibitemOpen
  \bibfield  {author} {\bibinfo {author} {\bibfnamefont {X.-R.}\ \bibnamefont
  {Jin}}, \bibinfo {author} {\bibfnamefont {Y.-H.}\ \bibnamefont {Lu}},
  \bibinfo {author} {\bibfnamefont {H.-Y.}\ \bibnamefont {Zheng}}, \bibinfo
  {author} {\bibfnamefont {Y.-P.}\ \bibnamefont {Lee}}, \bibinfo {author}
  {\bibfnamefont {J.-Y.}\ \bibnamefont {Rhee}}, \bibinfo {author}
  {\bibfnamefont {K.-W.}\ \bibnamefont {Kim}}, \ and\ \bibinfo {author}
  {\bibfnamefont {W.-H.}\ \bibnamefont {Jang}},\ }\bibfield  {title} {\enquote
  {\bibinfo {title} {Plasmonic electromagnetically-induced transparency in
  metamaterial based on second-order plasmonic resonance},}\ }\href {\doibase
  10.1016/j.optcom.2011.05.052} {\bibfield  {journal} {\bibinfo  {journal}
  {Opt. Commun.}\ }\textbf {\bibinfo {volume} {284}},\ \bibinfo {pages} {4766}
  (\bibinfo {year} {2011})}\BibitemShut {NoStop}%
\bibitem [{\citenamefont {Tassin}\ \emph {et~al.}(2009)\citenamefont {Tassin},
  \citenamefont {Zhang}, \citenamefont {Koschny}, \citenamefont {Economou},\
  and\ \citenamefont {Soukoulis}}]{tassin2009PRL}%
  \BibitemOpen
  \bibfield  {author} {\bibinfo {author} {\bibfnamefont {P.}~\bibnamefont
  {Tassin}}, \bibinfo {author} {\bibfnamefont {L.}~\bibnamefont {Zhang}},
  \bibinfo {author} {\bibfnamefont {Th.}\ \bibnamefont {Koschny}}, \bibinfo
  {author} {\bibfnamefont {E.~N.}\ \bibnamefont {Economou}}, \ and\ \bibinfo
  {author} {\bibfnamefont {C.~M.}\ \bibnamefont {Soukoulis}},\ }\bibfield
  {title} {\enquote {\bibinfo {title} {Low-loss metamaterials based on
  classical electromagnetically induced transparency},}\ }\href {\doibase
  10.1103/PhysRevLett.102.053901} {\bibfield  {journal} {\bibinfo  {journal}
  {Phys. Rev. Lett.}\ }\textbf {\bibinfo {volume} {102}},\ \bibinfo {pages}
  {053901} (\bibinfo {year} {2009})}\BibitemShut {NoStop}%
\bibitem [{\citenamefont {Xu}\ \emph {et~al.}(2010)\citenamefont {Xu},
  \citenamefont {Lu}, \citenamefont {Lee},\ and\ \citenamefont
  {Ham}}]{xu2010OE}%
  \BibitemOpen
  \bibfield  {author} {\bibinfo {author} {\bibfnamefont {H.}~\bibnamefont
  {Xu}}, \bibinfo {author} {\bibfnamefont {Y.-H.}\ \bibnamefont {Lu}}, \bibinfo
  {author} {\bibfnamefont {Y.-P.}\ \bibnamefont {Lee}}, \ and\ \bibinfo
  {author} {\bibfnamefont {B.-S.}\ \bibnamefont {Ham}},\ }\bibfield  {title}
  {\enquote {\bibinfo {title} {Studies of electromagnetically induced
  transparency in metamaterials},}\ }\href {\doibase 10.1364/OE.18.017736}
  {\bibfield  {journal} {\bibinfo  {journal} {Opt. Express}\ }\textbf {\bibinfo
  {volume} {18}},\ \bibinfo {pages} {17736} (\bibinfo {year}
  {2010})}\BibitemShut {NoStop}%
\bibitem [{\citenamefont {Huang}\ \emph {et~al.}(2022)\citenamefont {Huang},
  \citenamefont {Lin}, \citenamefont {Yao}, \citenamefont {Xia}, \citenamefont
  {Yin},\ and\ \citenamefont {Ai}}]{Huang2022AdP}%
  \BibitemOpen
  \bibfield  {author} {\bibinfo {author} {\bibfnamefont {H.-B.}\ \bibnamefont
  {Huang}}, \bibinfo {author} {\bibfnamefont {J.-J.}\ \bibnamefont {Lin}},
  \bibinfo {author} {\bibfnamefont {Y.-X.}\ \bibnamefont {Yao}}, \bibinfo
  {author} {\bibfnamefont {K.-Y.}\ \bibnamefont {Xia}}, \bibinfo {author}
  {\bibfnamefont {Z.-Q.}\ \bibnamefont {Yin}}, \ and\ \bibinfo {author}
  {\bibfnamefont {Q.}~\bibnamefont {Ai}},\ }\bibfield  {title} {\enquote
  {\bibinfo {title} {Optical nonreciprocity in rotating diamond with
  nitrogen-vacancy color centers},}\ }\href {\doibase 10.1002/andp.202200157}
  {\bibfield  {journal} {\bibinfo  {journal} {Ann. Phys. (Berlin)}\ }\textbf
  {\bibinfo {volume} {534}},\ \bibinfo {pages} {2200157} (\bibinfo {year}
  {2022})}\BibitemShut {NoStop}%
\bibitem [{\citenamefont {Boller}\ \emph {et~al.}(1991)\citenamefont {Boller},
  \citenamefont {Imamo\ifmmode~\breve{g}\else \u{g}\fi{}lu},\ and\
  \citenamefont {Harris}}]{boller1991PRL}%
  \BibitemOpen
  \bibfield  {author} {\bibinfo {author} {\bibfnamefont {K.-J.}\ \bibnamefont
  {Boller}}, \bibinfo {author} {\bibfnamefont {A.}~\bibnamefont
  {Imamo\ifmmode~\breve{g}\else \u{g}\fi{}lu}}, \ and\ \bibinfo {author}
  {\bibfnamefont {S.~E.}\ \bibnamefont {Harris}},\ }\bibfield  {title}
  {\enquote {\bibinfo {title} {Observation of electromagnetically induced
  transparency},}\ }\href {\doibase 10.1103/PhysRevLett.66.2593} {\bibfield
  {journal} {\bibinfo  {journal} {Phys. Rev. Lett.}\ }\textbf {\bibinfo
  {volume} {66}},\ \bibinfo {pages} {2593} (\bibinfo {year}
  {1991})}\BibitemShut {NoStop}%
\bibitem [{\citenamefont {Mukamel}(1999)}]{Mukamel1999}%
  \BibitemOpen
  \bibfield  {author} {\bibinfo {author} {\bibfnamefont {S.}~\bibnamefont
  {Mukamel}},\ }\href@noop {} {\emph {\bibinfo {title} {Principles of Nonlinear
  Optics and Spectroscopy}}}\ (\bibinfo  {publisher} {Oxford University Press,
  New York},\ \bibinfo {year} {1999})\BibitemShut {NoStop}%
\bibitem [{\citenamefont {Schlau-Cohen}\ \emph {et~al.}(2011)\citenamefont
  {Schlau-Cohen}, \citenamefont {Ishizaki},\ and\ \citenamefont
  {Fleming}}]{schlau2011CP}%
  \BibitemOpen
  \bibfield  {author} {\bibinfo {author} {\bibfnamefont {G.~S.}\ \bibnamefont
  {Schlau-Cohen}}, \bibinfo {author} {\bibfnamefont {A.}~\bibnamefont
  {Ishizaki}}, \ and\ \bibinfo {author} {\bibfnamefont {G.~R.}\ \bibnamefont
  {Fleming}},\ }\bibfield  {title} {\enquote {\bibinfo {title} {Two-dimensional
  electronic spectroscopy and photosynthesis: Fundamentals and applications to
  photosynthetic light-harvesting},}\ }\href {\doibase
  10.1016/j.chemphys.2011.04.025} {\bibfield  {journal} {\bibinfo  {journal}
  {Chem. Phys.}\ }\textbf {\bibinfo {volume} {386}},\ \bibinfo {pages} {1}
  (\bibinfo {year} {2011})}\BibitemShut {NoStop}%
\bibitem [{\citenamefont {Fuller}\ and\ \citenamefont
  {Ogilvie}(2015)}]{fuller2015ARPC}%
  \BibitemOpen
  \bibfield  {author} {\bibinfo {author} {\bibfnamefont {F.~D.}\ \bibnamefont
  {Fuller}}\ and\ \bibinfo {author} {\bibfnamefont {J.~P.}\ \bibnamefont
  {Ogilvie}},\ }\bibfield  {title} {\enquote {\bibinfo {title} {Experimental
  implementations of two-dimensional fourier transform electronic
  spectroscopy},}\ }\href {\doibase 10.1146/annurev-physchem-040513-103623}
  {\bibfield  {journal} {\bibinfo  {journal} {Annu. Rev. Phys. Chem.}\ }\textbf
  {\bibinfo {volume} {66}},\ \bibinfo {pages} {667} (\bibinfo {year}
  {2015})}\BibitemShut {NoStop}%
\bibitem [{\citenamefont {Zhu}\ \emph {et~al.}(2020)\citenamefont {Zhu},
  \citenamefont {Zou}, \citenamefont {Wang}, \citenamefont {Chen},\ and\ \citenamefont
  {Weng}}]{Zhu2020jpca}%
  \BibitemOpen
  \bibfield  {author} {\bibinfo {author} {\bibfnamefont {R.-D.}\ \bibnamefont
  {Zhu}}, \bibinfo {author} {\bibfnamefont {J.-D.}\ \bibnamefont {Zou}},
  \bibinfo {author} {\bibfnamefont {Z.}~\bibnamefont {Wang}}, \bibinfo {author} {\bibfnamefont {H.-L.}~\bibnamefont {Chen}},\ and\ \bibinfo
  {author} {\bibfnamefont {Y.-X.}\ \bibnamefont {Weng}},\ }\bibfield  {title}
  {\enquote {\bibinfo {title} {Electronic State-Resolved Multimode-Coupled Vibrational Wavepackets in Oxazine 720 by Two-Dimensional Electronic Spectroscopy},}\ }\href {\doibase
  10.1021/acs.jpca.0c06559} {\bibfield  {journal} {\bibinfo  {journal} {J. Phys. Chem. A}\ }\textbf {\bibinfo {volume} {124}},\ \bibinfo {pages}
  {9333} (\bibinfo {year} {2020})}\BibitemShut {NoStop}%
\bibitem [{\citenamefont {Sun}\ \emph {et~al.}(2023)\citenamefont {Sun},
  \citenamefont {Yao}, \citenamefont {Ai},\ and\ \citenamefont
  {Cheng}}]{sun2023AQT}%
  \BibitemOpen
  \bibfield  {author} {\bibinfo {author} {\bibfnamefont {Z.-H.}\ \bibnamefont
  {Sun}}, \bibinfo {author} {\bibfnamefont {Y.-X.}\ \bibnamefont {Yao}},
  \bibinfo {author} {\bibfnamefont {Q.}~\bibnamefont {Ai}}, \ and\ \bibinfo
  {author} {\bibfnamefont {Y.-C.}\ \bibnamefont {Cheng}},\ }\bibfield  {title}
  {\enquote {\bibinfo {title} {Theory of center-line slope in {2D} electronic
  spectroscopy with static disorder},}\ }\href {\doibase
  10.1002/qute.202300163} {\bibfield  {journal} {\bibinfo  {journal} {Adv.
  Quantum Technol.}\ }\textbf {\bibinfo {volume} {6}},\ \bibinfo {pages}
  {2300163} (\bibinfo {year} {2023})}\BibitemShut {NoStop}%
\bibitem [{\citenamefont {Ito}\ \emph {et~al.}(2014)\citenamefont {Ito},
  \citenamefont {Hasegawa},\ and\ \citenamefont {Tanimura}}]{Ito2014jcp}%
  \BibitemOpen
  \bibfield  {author} {\bibinfo {author} {\bibfnamefont {H.}\ \bibnamefont
  {Ito}}, \bibinfo {author} {\bibfnamefont {T.}\ \bibnamefont {Hasegawa}}, \
  and\ \bibinfo {author} {\bibfnamefont {Y.}\ \bibnamefont {Tanimura}},\
  }\bibfield  {title} {\enquote {\bibinfo {title} {Calculating two-dimensional THz-Raman-THz and Raman-THz-THz signals for various molecular liquids: The samplers},}\ }\href {\doibase
  10.1063/1.4895908} {\bibfield  {journal} {\bibinfo  {journal} {J.
  Chem. Phys.}\ }\textbf {\bibinfo {volume} {141}},\ \bibinfo {pages}
  {124503} (\bibinfo {year} {2014})}\BibitemShut {NoStop}%
\bibitem [{\citenamefont {Liu}\ \emph {et~al.}(2020)\citenamefont {Liu},
  \citenamefont {Sizhuk},\ and\ \citenamefont {Dorfman}}]{liu2020JPCL}%
  \BibitemOpen
  \bibfield  {author} {\bibinfo {author} {\bibfnamefont {G.-Y.}\ \bibnamefont
  {Liu}}, \bibinfo {author} {\bibfnamefont {A.~S.}\ \bibnamefont {Sizhuk}}, \
  and\ \bibinfo {author} {\bibfnamefont {K.~E.}\ \bibnamefont {Dorfman}},\
  }\bibfield  {title} {\enquote {\bibinfo {title} {Selective elimination of
  homogeneous broadening by multidimensional spectroscopy in the
  electromagnetically induced transparency regime},}\ }\href {\doibase
  10.1021/acs.jpclett.0c01481} {\bibfield  {journal} {\bibinfo  {journal} {J.
  Phys. Chem. Lett.}\ }\textbf {\bibinfo {volume} {11}},\ \bibinfo {pages}
  {5504} (\bibinfo {year} {2020})}\BibitemShut {NoStop}%
\bibitem [{\citenamefont {Nagata}\ \emph {et~al.}(2007)\citenamefont {Nagata},
  \citenamefont {Tanimura},\ and\ \citenamefont {Muckamel}}]{Nagata2007jcp}%
  \BibitemOpen
  \bibfield  {author} {\bibinfo {author} {\bibfnamefont {Y.}\ \bibnamefont
  {Nagata}}, \bibinfo {author} {\bibfnamefont {Y.}\ \bibnamefont {Tanimura}}, \
  and\ \bibinfo {author} {\bibfnamefont {S.}\ \bibnamefont {Muckamel}},\
  }\bibfield  {title} {\enquote {\bibinfo {title} {Two-dimensional infrared surface spectroscopy for CO on Cu(100): Detection of intermolecular coupling of adsorbates},}\ }\href {\doibase
  10.1063/1.2727445} {\bibfield  {journal} {\bibinfo  {journal} {J.
  Chem. Phys.}\ }\textbf {\bibinfo {volume} {126}},\ \bibinfo {pages}
  {204703} (\bibinfo {year} {2007})}\BibitemShut {NoStop}%
\bibitem [{\citenamefont {Seiler}\ \emph {et~al.}(2018)\citenamefont {Seiler},
  \citenamefont {Palato}, \citenamefont {Sonnichsen}, \citenamefont {Baker},\
  and\ \citenamefont {Kambhampati}}]{seiler2018NL}%
  \BibitemOpen
  \bibfield  {author} {\bibinfo {author} {\bibfnamefont {H.}~\bibnamefont
  {Seiler}}, \bibinfo {author} {\bibfnamefont {S.}~\bibnamefont {Palato}},
  \bibinfo {author} {\bibfnamefont {C.}~\bibnamefont {Sonnichsen}}, \bibinfo
  {author} {\bibfnamefont {H.}~\bibnamefont {Baker}}, \ and\ \bibinfo {author}
  {\bibfnamefont {P.}~\bibnamefont {Kambhampati}},\ }\bibfield  {title}
  {\enquote {\bibinfo {title} {Seeing multiexcitons through sample
  inhomogeneity: Band-edge biexciton structure in cdse nanocrystals revealed by
  two-dimensional electronic spectroscopy},}\ }\href {\doibase
  10.1021/acs.nanolett.8b00470} {\bibfield  {journal} {\bibinfo  {journal}
  {Nano Lett.}\ }\textbf {\bibinfo {volume} {18}},\ \bibinfo {pages} {2999.}
  (\bibinfo {year} {2018})}\BibitemShut {NoStop}%
\bibitem [{\citenamefont {Brosseau}\ \emph {et~al.}(2020)\citenamefont
  {Brosseau}, \citenamefont {Palato}, \citenamefont {Seiler}, \citenamefont
  {Baker},\ and\ \citenamefont {Kambhampati}}]{brosseau2020JCP}%
  \BibitemOpen
  \bibfield  {author} {\bibinfo {author} {\bibfnamefont {P.}~\bibnamefont
  {Brosseau}}, \bibinfo {author} {\bibfnamefont {S.}~\bibnamefont {Palato}},
  \bibinfo {author} {\bibfnamefont {H.}~\bibnamefont {Seiler}}, \bibinfo
  {author} {\bibfnamefont {H.}~\bibnamefont {Baker}}, \ and\ \bibinfo {author}
  {\bibfnamefont {P.}~\bibnamefont {Kambhampati}},\ }\bibfield  {title}
  {\enquote {\bibinfo {title} {Fifth-order two-quantum absorptive
  two-dimensional electronic spectroscopy of {CdSe} quantum dots},}\ }\href
  {\doibase 10.1063/5.0021381} {\bibfield  {journal} {\bibinfo  {journal} {J.
  Chem Phys.}\ }\textbf {\bibinfo {volume} {153}},\ \bibinfo {pages} {234703}
  (\bibinfo {year} {2020})}\BibitemShut {NoStop}%
\bibitem [{\citenamefont {Moody}\ \emph {et~al.}(2015)\citenamefont {Moody},
  \citenamefont {Kavir~Dass}, \citenamefont {Hao}, \citenamefont {Chen},
  \citenamefont {Li}, \citenamefont {Singh}, \citenamefont {Tran},
  \citenamefont {Clark}, \citenamefont {Xu}, \citenamefont {Bergh{\"a}user},
  \citenamefont {Malic}, \citenamefont {Knorr},\ and\ \citenamefont
  {Li}}]{moody2015NC}%
  \BibitemOpen
  \bibfield  {author} {\bibinfo {author} {\bibfnamefont {G.}~\bibnamefont
  {Moody}}, \bibinfo {author} {\bibfnamefont {C.}~\bibnamefont {Kavir~Dass}},
  \bibinfo {author} {\bibfnamefont {K.}~\bibnamefont {Hao}}, \bibinfo {author}
  {\bibfnamefont {C.-H.}\ \bibnamefont {Chen}}, \bibinfo {author}
  {\bibfnamefont {L.-J.}\ \bibnamefont {Li}}, \bibinfo {author} {\bibfnamefont
  {A.}~\bibnamefont {Singh}}, \bibinfo {author} {\bibfnamefont
  {K.}~\bibnamefont {Tran}}, \bibinfo {author} {\bibfnamefont {G.}~\bibnamefont
  {Clark}}, \bibinfo {author} {\bibfnamefont {X.-D.}\ \bibnamefont {Xu}},
  \bibinfo {author} {\bibfnamefont {G.}~\bibnamefont {Bergh{\"a}user}},
  \bibinfo {author} {\bibfnamefont {E.}~\bibnamefont {Malic}}, \bibinfo
  {author} {\bibfnamefont {A.}~\bibnamefont {Knorr}}, \ and\ \bibinfo {author}
  {\bibfnamefont {X.-Q.}\ \bibnamefont {Li}},\ }\bibfield  {title} {\enquote
  {\bibinfo {title} {Intrinsic homogeneous linewidth and broadening mechanisms
  of excitons in monolayer transition metal dichalcogenides},}\ }\href
  {\doibase 10.1038/ncomms9315} {\bibfield  {journal} {\bibinfo  {journal}
  {Nat. Commun.}\ }\textbf {\bibinfo {volume} {6}},\ \bibinfo {pages} {8315}
  (\bibinfo {year} {2015})}\BibitemShut {NoStop}%
\bibitem [{\citenamefont {Guo}\ \emph {et~al.}(2019)\citenamefont {Guo},
  \citenamefont {Wu}, \citenamefont {Cao}, \citenamefont {Monahan},
  \citenamefont {Lee}, \citenamefont {Louie},\ and\ \citenamefont
  {Fleming}}]{guo2019NP}%
  \BibitemOpen
  \bibfield  {author} {\bibinfo {author} {\bibfnamefont {L.}~\bibnamefont
  {Guo}}, \bibinfo {author} {\bibfnamefont {M.}~\bibnamefont {Wu}}, \bibinfo
  {author} {\bibfnamefont {T.}~\bibnamefont {Cao}}, \bibinfo {author}
  {\bibfnamefont {D.~M.}\ \bibnamefont {Monahan}}, \bibinfo {author}
  {\bibfnamefont {Y.-H.}\ \bibnamefont {Lee}}, \bibinfo {author} {\bibfnamefont
  {S.~G.}\ \bibnamefont {Louie}}, \ and\ \bibinfo {author} {\bibfnamefont
  {G.~R.}\ \bibnamefont {Fleming}},\ }\bibfield  {title} {\enquote {\bibinfo
  {title} {Exchange-driven intravalley mixing of excitons in monolayer
  transition metal dichalcogenides},}\ }\href {\doibase
  10.1038/s41567-018-0362-y} {\bibfield  {journal} {\bibinfo  {journal} {Nat.
  Phys.}\ }\textbf {\bibinfo {volume} {15}},\ \bibinfo {pages} {228} (\bibinfo
  {year} {2019})}\BibitemShut {NoStop}%
\bibitem [{\citenamefont {Jha}\ \emph {et~al.}(2017)\citenamefont {Jha},
  \citenamefont {Duan}, \citenamefont {Tiwari}, \citenamefont {Nayak},
  \citenamefont {Snaith}, \citenamefont {Thorwart},\ and\ \citenamefont
  {Miller}}]{jha2017AP}%
  \BibitemOpen
  \bibfield  {author} {\bibinfo {author} {\bibfnamefont {A.}~\bibnamefont
  {Jha}}, \bibinfo {author} {\bibfnamefont {H.-G.}\ \bibnamefont {Duan}},
  \bibinfo {author} {\bibfnamefont {V.}~\bibnamefont {Tiwari}}, \bibinfo
  {author} {\bibfnamefont {P.~K.}\ \bibnamefont {Nayak}}, \bibinfo {author}
  {\bibfnamefont {H.~J.}\ \bibnamefont {Snaith}}, \bibinfo {author}
  {\bibfnamefont {M.}~\bibnamefont {Thorwart}}, \ and\ \bibinfo {author}
  {\bibfnamefont {R.~J.~D.}\ \bibnamefont {Miller}},\ }\bibfield  {title}
  {\enquote {\bibinfo {title} {Direct observation of ultrafast exciton
  dissociation in lead iodide perovskite by {2D} electronic spectroscopy},}\
  }\href {\doibase 10.1021/acsphotonics.7b01025} {\bibfield  {journal}
  {\bibinfo  {journal} {Acs Photonics}\ }\textbf {\bibinfo {volume} {5}},\
  \bibinfo {pages} {852} (\bibinfo {year} {2017})}\BibitemShut {NoStop}%
\bibitem [{\citenamefont {Monahan}\ \emph {et~al.}(2017)\citenamefont
  {Monahan}, \citenamefont {Guo}, \citenamefont {Lin}, \citenamefont {Dou},
  \citenamefont {Yang},\ and\ \citenamefont {Fleming}}]{monahan2017JPCL}%
  \BibitemOpen
  \bibfield  {author} {\bibinfo {author} {\bibfnamefont {D.~M.}\ \bibnamefont
  {Monahan}}, \bibinfo {author} {\bibfnamefont {L.}~\bibnamefont {Guo}},
  \bibinfo {author} {\bibfnamefont {J.}~\bibnamefont {Lin}}, \bibinfo {author}
  {\bibfnamefont {L.-T.}\ \bibnamefont {Dou}}, \bibinfo {author} {\bibfnamefont
  {P.-D.}\ \bibnamefont {Yang}}, \ and\ \bibinfo {author} {\bibfnamefont
  {G.~R.}\ \bibnamefont {Fleming}},\ }\bibfield  {title} {\enquote {\bibinfo
  {title} {{Room-temperature coherent optical phonon in 2D electronic spectra
  of CH$_3$NH$_3$PbI$_3$ perovskite as a possible cooling bottleneck}},}\
  }\href {\doibase 10.1021/acs.jpclett.7b01357} {\bibfield  {journal} {\bibinfo
   {journal} {J. Phys. Chem. Lett.}\ }\textbf {\bibinfo {volume} {8}},\
  \bibinfo {pages} {3211} (\bibinfo {year} {2017})}\BibitemShut {NoStop}%
\bibitem [{\citenamefont {Seiler}\ \emph {et~al.}(2019)\citenamefont {Seiler},
  \citenamefont {Palato}, \citenamefont {Sonnichsen}, \citenamefont {Baker},
  \citenamefont {Socie}, \citenamefont {Strandell},\ and\ \citenamefont
  {Kambhampati}}]{seiler2019NC}%
  \BibitemOpen
  \bibfield  {author} {\bibinfo {author} {\bibfnamefont {H.}~\bibnamefont
  {Seiler}}, \bibinfo {author} {\bibfnamefont {S.}~\bibnamefont {Palato}},
  \bibinfo {author} {\bibfnamefont {C.}~\bibnamefont {Sonnichsen}}, \bibinfo
  {author} {\bibfnamefont {H.}~\bibnamefont {Baker}}, \bibinfo {author}
  {\bibfnamefont {E.}~\bibnamefont {Socie}}, \bibinfo {author} {\bibfnamefont
  {D.~P.}\ \bibnamefont {Strandell}}, \ and\ \bibinfo {author} {\bibfnamefont
  {P.}~\bibnamefont {Kambhampati}},\ }\bibfield  {title} {\enquote {\bibinfo
  {title} {Two-dimensional electronic spectroscopy reveals liquid-like
  lineshape dynamics in CsPbI$_3$ perovskite nanocrystals},}\ }\href {\doibase
  10.1038/s41467-019-12830-1} {\bibfield  {journal} {\bibinfo  {journal} {Nat.
  Commun.}\ }\textbf {\bibinfo {volume} {10}},\ \bibinfo {pages} {4962}
  (\bibinfo {year} {2019})}\BibitemShut {NoStop}%
\bibitem [{\citenamefont {Calhoun}\ \emph {et~al.}(2009)\citenamefont
  {Calhoun}, \citenamefont {Ginsberg}, \citenamefont {Schlau-Cohen},
  \citenamefont {Cheng}, \citenamefont {Ballottari}, \citenamefont {Bassi},\
  and\ \citenamefont {Fleming}}]{calhoun2009JPCB}%
  \BibitemOpen
  \bibfield  {author} {\bibinfo {author} {\bibfnamefont {T.~R.}\ \bibnamefont
  {Calhoun}}, \bibinfo {author} {\bibfnamefont {N.~S.}\ \bibnamefont
  {Ginsberg}}, \bibinfo {author} {\bibfnamefont {G.~S.}\ \bibnamefont
  {Schlau-Cohen}}, \bibinfo {author} {\bibfnamefont {Y.-C.}\ \bibnamefont
  {Cheng}}, \bibinfo {author} {\bibfnamefont {M.}~\bibnamefont {Ballottari}},
  \bibinfo {author} {\bibfnamefont {R.}~\bibnamefont {Bassi}}, \ and\ \bibinfo
  {author} {\bibfnamefont {G.~R.}\ \bibnamefont {Fleming}},\ }\bibfield
  {title} {\enquote {\bibinfo {title} {Quantum coherence enabled determination
  of the energy landscape in light-harvesting complex {II}},}\ }\href {\doibase
  10.1021/jp908300c} {\bibfield  {journal} {\bibinfo  {journal} {J. Phys. Chem.
  B}\ }\textbf {\bibinfo {volume} {113}},\ \bibinfo {pages} {16291} (\bibinfo
  {year} {2009})}\BibitemShut {NoStop}%
\bibitem [{\citenamefont {Fuller}\ \emph {et~al.}(2014)\citenamefont {Fuller},
  \citenamefont {Pan}, \citenamefont {Gelzinis}, \citenamefont {Butkus},
  \citenamefont {Senlik}, \citenamefont {Wilcox}, \citenamefont {Yocum},
  \citenamefont {Valkunas}, \citenamefont {Abramavicius},\ and\ \citenamefont
  {Ogilvie}}]{fuller2014NC}%
  \BibitemOpen
  \bibfield  {author} {\bibinfo {author} {\bibfnamefont {F.~D.}\ \bibnamefont
  {Fuller}}, \bibinfo {author} {\bibfnamefont {J.}~\bibnamefont {Pan}},
  \bibinfo {author} {\bibfnamefont {A.}~\bibnamefont {Gelzinis}}, \bibinfo
  {author} {\bibfnamefont {V.}~\bibnamefont {Butkus}}, \bibinfo {author}
  {\bibfnamefont {S.~S.}\ \bibnamefont {Senlik}}, \bibinfo {author}
  {\bibfnamefont {D.~E.}\ \bibnamefont {Wilcox}}, \bibinfo {author}
  {\bibfnamefont {C.~F.}\ \bibnamefont {Yocum}}, \bibinfo {author}
  {\bibfnamefont {L.}~\bibnamefont {Valkunas}}, \bibinfo {author}
  {\bibfnamefont {D.}~\bibnamefont {Abramavicius}}, \ and\ \bibinfo {author}
  {\bibfnamefont {J.~P.}\ \bibnamefont {Ogilvie}},\ }\bibfield  {title}
  {\enquote {\bibinfo {title} {Vibronic coherence in oxygenic
  photosynthesis},}\ }\href {\doibase 10.1038/NCHEM.2005} {\bibfield  {journal}
  {\bibinfo  {journal} {Nat. Chem.}\ }\textbf {\bibinfo {volume} {6}},\
  \bibinfo {pages} {706} (\bibinfo {year} {2014})}\BibitemShut {NoStop}%
\bibitem [{\citenamefont {Zhu}\ \emph {et~al.}(2024)\citenamefont {Zhu},
  \citenamefont {Li}, \citenamefont {Zhen}, \citenamefont {Zou},
  \citenamefont {Liao}, \citenamefont {Wang}, \citenamefont {Wang}, \citenamefont {Chen}, \citenamefont {Qin} \ and\ \citenamefont {Weng}}]{Zhu2024nc}%
  \BibitemOpen
  \bibfield  {author} {\bibinfo {author} {\bibfnamefont {R.-D.}\ \bibnamefont
  {Zhu}}, \bibinfo {author} {\bibfnamefont {W.-J.}\ \bibnamefont {Li}},
  \bibinfo {author} {\bibfnamefont {Z.-H.}~\bibnamefont {Zhen}}, \bibinfo
  {author} {\bibfnamefont {J.-D.}~\bibnamefont {Zou}}, \bibinfo {author}
  {\bibfnamefont {G.-H.}\ \bibnamefont {Liao}}, \bibinfo {author}
  {\bibfnamefont {J.-Y.}\ \bibnamefont {Wang}}, \bibinfo {author}
  {\bibfnamefont {Z.}\ \bibnamefont {Wang}}, \bibinfo {author}
  {\bibfnamefont {H.-L.}\ \bibnamefont {Chen}}, \bibinfo {author}
  {\bibfnamefont {S.}\ \bibnamefont {Qin}}, \ and\ \bibinfo {author}
  {\bibfnamefont {Y.-X.}~\bibnamefont {Weng}},\ }\bibfield  {title} {\enquote
  {\bibinfo {title} {Quantum phase synchronization via exciton-vibrational energy dissipation sustains long-lived coherence in photosynthetic antennas},}\ }\href {\doibase
  10.1038/s41467-024-47560-6} {\bibfield  {journal} {\bibinfo  {journal} {Nat. Commun.
  }\ }\textbf {\bibinfo {volume} {15}},\ \bibinfo {pages} {3171} (\bibinfo
  {year} {2024})}\BibitemShut {NoStop}%
\bibitem [{\citenamefont {Romero}\ \emph {et~al.}(2014)\citenamefont {Romero},
  \citenamefont {Augulis}, \citenamefont {Novoderezhkin}, \citenamefont
  {Ferretti}, \citenamefont {Thieme}, \citenamefont {Zigmantas},\ and\
  \citenamefont {van Grondelle}}]{romero2014NP}%
  \BibitemOpen
  \bibfield  {author} {\bibinfo {author} {\bibfnamefont {E.}~\bibnamefont
  {Romero}}, \bibinfo {author} {\bibfnamefont {R.}~\bibnamefont {Augulis}},
  \bibinfo {author} {\bibfnamefont {V.~I.}\ \bibnamefont {Novoderezhkin}},
  \bibinfo {author} {\bibfnamefont {M.}~\bibnamefont {Ferretti}}, \bibinfo
  {author} {\bibfnamefont {J.}~\bibnamefont {Thieme}}, \bibinfo {author}
  {\bibfnamefont {D.}~\bibnamefont {Zigmantas}}, \ and\ \bibinfo {author}
  {\bibfnamefont {R.}~\bibnamefont {van Grondelle}},\ }\bibfield  {title}
  {\enquote {\bibinfo {title} {Quantum coherence in photosynthesis for
  efficient solar-energy conversion},}\ }\href {\doibase 10.1038/NPHYS3017}
  {\bibfield  {journal} {\bibinfo  {journal} {Nat. Phys.}\ }\textbf {\bibinfo
  {volume} {10}},\ \bibinfo {pages} {676} (\bibinfo {year} {2014})}\BibitemShut
  {NoStop}%
\bibitem [{\citenamefont {Policht}\ \emph {et~al.}(2022)\citenamefont {Policht},
  \citenamefont {Niedringhaus}, \citenamefont {Willow}, \citenamefont {Laible},
  \citenamefont {Bocian}, \citenamefont {Kirmaier}, \citenamefont {Holten}, \citenamefont {Man{\v{c}}al},\ and\ \citenamefont
  {Ogilvie}}]{Policht2022sa}%
  \BibitemOpen
  \bibfield  {author} {\bibinfo {author} {\bibfnamefont {V.~R.}\ \bibnamefont
  {Policht}}, \bibinfo {author} {\bibfnamefont {A.}\ \bibnamefont {Niedringhaus}},
  \bibinfo {author} {\bibfnamefont {R.}\ \bibnamefont {Willow}}, \bibinfo
  {author} {\bibfnamefont {P.~D.}~\bibnamefont {Laible}}, \bibinfo {author}
  {\bibfnamefont {D.~F.}~\bibnamefont {Bocian}}, \bibinfo {author} {\bibfnamefont
  {C.}~\bibnamefont {Kirmaier}}, \bibinfo {author} {\bibfnamefont
  {D.}~\bibnamefont {Holten}}, \bibinfo {author} {\bibfnamefont
  {T.}~\bibnamefont {Man{\v{c}}al}},\ and\ \bibinfo {author} {\bibfnamefont {J.~P.}\
  \bibnamefont {Ogilvie}},\ }\bibfield  {title} {\enquote {\bibinfo {title} {Hidden vibronic and excitonic structure and vibronic coherence transfer in the bacterial reaction center},}\ }\href {\doibase
  10.1126/sciadv.abk0953} {\bibfield  {journal} {\bibinfo  {journal} {Sci. Adv.}\ }\textbf {\bibinfo {volume} {8}},\ \bibinfo {pages} {eabk0953}
  (\bibinfo {year} {2022})}\BibitemShut {NoStop}%
\bibitem [{\citenamefont {Ma}\ \emph {et~al.}(2017)\citenamefont {Ma},
  \citenamefont {Yu}, \citenamefont {Hendrikx}, \citenamefont {Wang~Otomo},\
  and\ \citenamefont {van Grondelle}}]{ma2017JPCL}%
  \BibitemOpen
  \bibfield  {author} {\bibinfo {author} {\bibfnamefont {F.}~\bibnamefont
  {Ma}}, \bibinfo {author} {\bibfnamefont {L.-J.}\ \bibnamefont {Yu}}, \bibinfo
  {author} {\bibfnamefont {R.}~\bibnamefont {Hendrikx}}, \bibinfo {author}
  {\bibfnamefont {Z.~Y.}\ \bibnamefont {Wang~Otomo}}, \ and\ \bibinfo {author}
  {\bibfnamefont {R.}~\bibnamefont {van Grondelle}},\ }\bibfield  {title}
  {\enquote {\bibinfo {title} {Excitonic and vibrational coherence in the
  excitation relaxation process of two {LH1} complexes as revealed by
  two-dimensional electronic spectroscopy},}\ }\href {\doibase
  10.1021/acs.jpclett.7b00824} {\bibfield  {journal} {\bibinfo  {journal} {J.
  Phys. Chem. Lett.}\ }\textbf {\bibinfo {volume} {8}},\ \bibinfo {pages}
  {2751} (\bibinfo {year} {2017})}\BibitemShut {NoStop}%
\bibitem [{\citenamefont {Thyrhaug}\ \emph {et~al.}(2018)\citenamefont
  {Thyrhaug}, \citenamefont {Tempelaar}, \citenamefont {Alcocer}, \citenamefont
  {{\v{Z}}{\'\i}dek}, \citenamefont {B{\'\i}na}, \citenamefont {Knoester},
  \citenamefont {Jansen},\ and\ \citenamefont {Zigmantas}}]{thyrhaug2018NC}%
  \BibitemOpen
  \bibfield  {author} {\bibinfo {author} {\bibfnamefont {E.}~\bibnamefont
  {Thyrhaug}}, \bibinfo {author} {\bibfnamefont {R.}~\bibnamefont {Tempelaar}},
  \bibinfo {author} {\bibfnamefont {M.~J.~P.}\ \bibnamefont {Alcocer}},
  \bibinfo {author} {\bibfnamefont {K.}~\bibnamefont {{\v{Z}}{\'\i}dek}},
  \bibinfo {author} {\bibfnamefont {D.}~\bibnamefont {B{\'\i}na}}, \bibinfo
  {author} {\bibfnamefont {J.}~\bibnamefont {Knoester}}, \bibinfo {author}
  {\bibfnamefont {T.~L.~C.}\ \bibnamefont {Jansen}}, \ and\ \bibinfo {author}
  {\bibfnamefont {D.}~\bibnamefont {Zigmantas}},\ }\bibfield  {title} {\enquote
  {\bibinfo {title} {Identification and characterization of diverse coherences
  in the {Fenna--Matthews--Olson} complex},}\ }\href {\doibase
  10.1038/s41557-018-0060-5} {\bibfield  {journal} {\bibinfo  {journal} {Nat.
  Chem.}\ }\textbf {\bibinfo {volume} {10}},\ \bibinfo {pages} {780} (\bibinfo
  {year} {2018})}\BibitemShut {NoStop}%
\bibitem [{\citenamefont {Silori}\ \emph {et~al.}(2023)\citenamefont {Silori},
  \citenamefont {Willow}, \citenamefont {Nguyen}, \citenamefont {Shen},
  \citenamefont {Song}, \citenamefont {Gisriel}, \citenamefont {Brudvig}, \citenamefont {Bryant},\ and\ \citenamefont
  {Ogilvie}}]{Silori2023jpcl}%
  \BibitemOpen
  \bibfield  {author} {\bibinfo {author} {\bibfnamefont {Y.}\ \bibnamefont
  {Silori}}, \bibinfo {author} {\bibfnamefont {R.}\ \bibnamefont {Willow}},
  \bibinfo {author} {\bibfnamefont {H.~H.}\ \bibnamefont {Nguyen}}, \bibinfo
  {author} {\bibfnamefont {G.-Z.}~\bibnamefont {Shen}}, \bibinfo {author}
  {\bibfnamefont {Y.}~\bibnamefont {Song}}, \bibinfo {author} {\bibfnamefont
  {C.~J.}~\bibnamefont {Gisriel}}, \bibinfo {author} {\bibfnamefont
  {G.~W.}~\bibnamefont {Brudvig}}, \bibinfo {author} {\bibfnamefont
  {D.~A.}~\bibnamefont {Bryant}},\ and\ \bibinfo {author} {\bibfnamefont {J.~P.}\
  \bibnamefont {Ogilvie}},\ }\bibfield  {title} {\enquote {\bibinfo {title} {Two-Dimensional Electronic Spectroscopy of the Far-Red-Light Photosystem II Reaction Center},}\ }\href {\doibase
  10.1021/acs.jpclett.3c02604} {\bibfield  {journal} {\bibinfo  {journal} {J. Phys. Chem. Lett.}\ }\textbf {\bibinfo {volume} {14}},\ \bibinfo {pages} {10300}
  (\bibinfo {year} {2023})}\BibitemShut {NoStop}%
\bibitem [{\citenamefont {Zhu}\ \emph {et~al.}(2022)\citenamefont {Zhu},
  \citenamefont {Ruan}, \citenamefont {Li}, \citenamefont {Leng},
  \citenamefont {Zou}, \citenamefont {Wang}, \citenamefont {Chen}, \citenamefont {Wang},\ and\ \citenamefont
  {Weng}}]{Zhu2022jcp}%
  \BibitemOpen
  \bibfield  {author} {\bibinfo {author} {\bibfnamefont {R.-D.}\ \bibnamefont
  {Zhu}}, \bibinfo {author} {\bibfnamefont {M.-X.}\ \bibnamefont {Ruan}},
  \bibinfo {author} {\bibfnamefont {H.}\ \bibnamefont {Li}}, \bibinfo
  {author} {\bibfnamefont {X.}~\bibnamefont {Leng}}, \bibinfo {author}
  {\bibfnamefont {J.-D.}~\bibnamefont {Zou}}, \bibinfo {author} {\bibfnamefont
  {J.-Y.}~\bibnamefont {Wang}}, \bibinfo {author} {\bibfnamefont
  {H.-L.}~\bibnamefont {Chen}}, \bibinfo {author} {\bibfnamefont
  {Z.}~\bibnamefont {Wang}},\ and\ \bibinfo {author} {\bibfnamefont {Y.-X.}\
  \bibnamefont {Weng}},\ }\bibfield  {title} {\enquote {\bibinfo {title} {Vibrational and vibronic coherences in the energy transfer process of light-harvesting complex II revealed by two-dimensional electronic spectroscopy},}\ }\href {\doibase
  10.1063/5.0082280} {\bibfield  {journal} {\bibinfo  {journal} {J. Chem. Phys.}\ }\textbf {\bibinfo {volume} {156}},\ \bibinfo {pages} {125101}
  (\bibinfo {year} {2022})}\BibitemShut {NoStop}%
\bibitem [{\citenamefont {Zhang}\ \emph {et~al.}(2021)\citenamefont {Zhang},
  \citenamefont {Borrelli},\ and\ \citenamefont
  {Tanimura}}]{Zhang2021jcp}%
  \BibitemOpen
  \bibfield  {author} {\bibinfo {author} {\bibfnamefont {J.-J.}\ \bibnamefont
  {Zhang}}, \bibinfo {author} {\bibfnamefont {R.}\ \bibnamefont {Borrelli}},\ and\ \bibinfo {author} {\bibfnamefont {Y.}\
  \bibnamefont {Tanimura}},\ }\bibfield  {title} {\enquote {\bibinfo {title} {Probing photoinduced proton coupled electron transfer process by means of two-dimensional resonant electronic-vibrational spectroscopy},}\ }\href {\doibase
  10.1063/5.0046755} {\bibfield  {journal} {\bibinfo  {journal} {J. Chem. Phys.}\ }\textbf {\bibinfo {volume} {154}},\ \bibinfo {pages} {144104}
  (\bibinfo {year} {2021})}\BibitemShut {NoStop}%
\bibitem [{\citenamefont {Deng}\ \emph {et~al.}(2024)\citenamefont {Deng},
  \citenamefont {Liu}, \citenamefont {Yao}, \citenamefont {Jin}, \citenamefont {Zhang}, \citenamefont {Song}, and\ \citenamefont {Ai}}]{Deng2024pra}%
  \BibitemOpen
  \bibfield  {author} {\bibinfo {author} {\bibfnamefont {R.-Q.}~\bibnamefont
  {Deng}}, \bibinfo {author} {\bibfnamefont {C.-G.}\ \bibnamefont {Liu}}, \bibinfo {author} {\bibfnamefont {Y.-X.}\ \bibnamefont {Yao}}, \bibinfo {author} {\bibfnamefont {J.-Y.-R.}\ \bibnamefont {Jin}}, \bibinfo {author} {\bibfnamefont {H.-Y.}\ \bibnamefont {Zhang}}, \bibinfo {author} {\bibfnamefont {Y.}\ \bibnamefont {Song}} and\ \bibinfo {author}
  {\bibfnamefont {Q.}~\bibnamefont {Ai}},\ }\bibfield  {title}
  {\enquote {\bibinfo {title} {Anomalously reduced homogeneous broadening of two-dimensional electronic spectroscopy at high temperature by detailed balance},}\ }\href {\doibase
  10.1103/PhysRevA.109.052801} {\bibfield  {journal} {\bibinfo  {journal} {Phys. Rev. A}\ }\textbf {\bibinfo {volume} {109}},\ \bibinfo {pages}
  {052801} (\bibinfo {year} {2024})}\BibitemShut {NoStop}%
\bibitem [{\citenamefont {Steck}(2024)\citenamefont {Steck}
   }]{Rb85}%
  \BibitemOpen
  \bibfield  {author} {\bibinfo {author}
  {\bibfnamefont {D.~A.}~\bibnamefont {Steck}},\ }\bibfield  {title}
  {\enquote {\bibinfo {title} {Rubidium 85 D Line Data},} }
  \href {\doibase
  } {\bibfield  {journal} {\bibinfo  {journal} {http://steck.us/alkalidata }\ }\textbf {\bibinfo {volume} { }}\ \bibinfo {pages} { }\hspace{-1em}(\bibinfo {year} {2024})}\BibitemShut {NoStop}%
\bibitem [{\citenamefont {Breuer}\ and\ \citenamefont
  {Petruccione}(2002)}]{Breuer2002}%
  \BibitemOpen
  \bibfield  {author} {\bibinfo {author} {\bibfnamefont {H.~P.}\ \bibnamefont
  {Breuer}}\ and\ \bibinfo {author} {\bibfnamefont {F.}~\bibnamefont
  {Petruccione}},\ }\href {\doibase 10.1093/acprof:oso/9780199213900.001.0001}
  {\emph {\bibinfo {title} {The Theory of Open Quantum Systems}}}\ (\bibinfo
  {publisher} {Oxford University Presss},\ \bibinfo {year} {2002})\BibitemShut
  {NoStop}%
\bibitem [{\citenamefont {Takagahara}\ \emph {et~al.}(1977)\citenamefont {Takagahara},
  \citenamefont {Hanamura},  and\ \citenamefont {Kubo}}]{Takagahara1977jpsj}%
  \BibitemOpen
  \bibfield  {author} {\bibinfo {author} {\bibfnamefont {T.}~\bibnamefont
  {Takagahara}}, \bibinfo {author} {\bibfnamefont {E.}\ \bibnamefont {Hanamura}} and\ \bibinfo {author}
  {\bibfnamefont {R.}~\bibnamefont {Kubo}},\ }\bibfield  {title}
  {\enquote {\bibinfo {title} {Stochastic Models of Intermediate State Interaction in Second Order Optical Processes -Stationary Response. II-},}\ }\href {\doibase
  10.1143/JPSJ.43.811} {\bibfield  {journal} {\bibinfo  {journal} {J. Phys. Soc. Jpn}\ }\textbf {\bibinfo {volume} {43}},\ \bibinfo {pages}
  {811} (\bibinfo {year} {1977})}\BibitemShut {NoStop}%
\bibitem [{\citenamefont {Tanimura}\ \emph {et~al.}(1989)\citenamefont {Tanimura},
  \citenamefont {Suzuki},  and\ \citenamefont {Kubo}}]{Tanimura1989jpsj}%
  \BibitemOpen
  \bibfield  {author} {\bibinfo {author} {\bibfnamefont {Y.}~\bibnamefont
  {Tanimura}}, \bibinfo {author} {\bibfnamefont {T.}\ \bibnamefont {Suzuki}} and\ \bibinfo {author}
  {\bibfnamefont {R.}~\bibnamefont {Kubo}},\ }\bibfield  {title}
  {\enquote {\bibinfo {title} {Second Order Optical Process of a Three-Level System in Contact with a Nearly Gaussian-Markoffian Noise Bath},}\ }\href {\doibase
  10.1143/JPSJ.58.1850} {\bibfield  {journal} {\bibinfo  {journal} {J. Phys. Soc. Jpn}\ }\textbf {\bibinfo {volume} {58}},\ \bibinfo {pages}
  {1850} (\bibinfo {year} {1989})}\BibitemShut {NoStop}%
\bibitem [{\citenamefont {Tanimura}(2006)\citenamefont {Tanimura}
   }]{Tanimura2006jpsj}%
  \BibitemOpen
  \bibfield  {author} {\bibinfo {author}
  {\bibfnamefont {Y.}~\bibnamefont {Tanimura}},\ }\bibfield  {title}
  {\enquote {\bibinfo {title} {Stochastic Liouville, Langevin, Fokker-Planck, and Master Equation Approaches to Quantum Dissipative Systems},}\ }\href {\doibase
  10.1143/JPSJ.75.082001} {\bibfield  {journal} {\bibinfo  {journal} {J. Phys. Soc. Jpn}\ }\textbf {\bibinfo {volume} {75}},\ \bibinfo {pages}
  {082001} (\bibinfo {year} {2006})}\BibitemShut {NoStop}%
\bibitem [{\citenamefont {Leggett}\ \emph {et~al.}(1987)\citenamefont {Leggett},
  \citenamefont {Chakravarty}, \citenamefont {Dorsey}, \citenamefont {Fisher}, \citenamefont {Garg}, and\ \citenamefont {Zwerger}}]{Leggett1987rmp}%
  \BibitemOpen
  \bibfield  {author} {\bibinfo {author} {\bibfnamefont {A.~J.}~\bibnamefont
  {Leggett}}, \bibinfo {author} {\bibfnamefont {S.}\ \bibnamefont {Chakravarty}}, \bibinfo {author} {\bibfnamefont {A.~T.}\ \bibnamefont {Dorsey}}, \bibinfo {author} {\bibfnamefont {Matthew P.~A.}\ \bibnamefont {Fisher}}, \bibinfo {author} {\bibfnamefont {A.}\ \bibnamefont {Garg}}, and\ \bibinfo {author}
  {\bibfnamefont {W.}~\bibnamefont {Zwerger}},\ }\bibfield  {title}
  {\enquote {\bibinfo {title} {Dynamics of the dissipative two-state system},}\ }\href {\doibase
  10.1103/RevModPhys.59.1} {\bibfield  {journal} {\bibinfo  {journal} {Rev. Mod. Phys.}\ }\textbf {\bibinfo {volume} {59}},\ \bibinfo {pages}
  {1} (\bibinfo {year} {1987})}\BibitemShut {NoStop}%
\bibitem [{\citenamefont {Koyanagi}\ \emph {et~al.}(2024)\citenamefont {Koyanagi},
   and\ \citenamefont {Tanimura}}]{Koyanagi2024jcp}%
  \BibitemOpen
  \bibfield  {author} {\bibinfo {author} {\bibfnamefont {S.}~\bibnamefont
  {Koyanagi}} and\ \bibinfo {author}
  {\bibfnamefont {Y.}~\bibnamefont {Tanimura}},\ }\bibfield  {title}
  {\enquote {\bibinfo {title} {Hierarchical equations of motion for multiple baths (HEOM-MB) and their
   application to Carnot cycle},}\ }\href {\doibase
  10.1063/5.0232073} {\bibfield  {journal} {\bibinfo  {journal} {J. Chem. Phys.}\ }\textbf {\bibinfo {volume} {161}},\ \bibinfo {pages}
  {162501} (\bibinfo {year} {2024})}\BibitemShut {NoStop}%
\bibitem [{\citenamefont {Tanimura}(2020)\citenamefont {Tanimura}
   }]{Tanimura2020jcp}%
  \BibitemOpen
  \bibfield  {author} {\bibinfo {author}
  {\bibfnamefont {Y.}~\bibnamefont {Tanimura}},\ }\bibfield  {title}
  {\enquote {\bibinfo {title} {Numerically ``exact{''} approach to open quantum dynamics: The
   hierarchical equations of motion (HEOM)},}\ }\href {\doibase
  10.1063/5.0011599} {\bibfield  {journal} {\bibinfo  {journal} {J. Chem. Phys.}\ }\textbf {\bibinfo {volume} {153}},\ \bibinfo {pages}
  {020901} (\bibinfo {year} {2020})}\BibitemShut {NoStop}%
\bibitem [{\citenamefont {Tanimura}(2014)\citenamefont {Tanimura}
   }]{Tanimura2014jcp}%
  \BibitemOpen
  \bibfield  {author} {\bibinfo {author}
  {\bibfnamefont {Y.}~\bibnamefont {Tanimura}},\ }\bibfield  {title}
  {\enquote {\bibinfo {title} {Reduced hierarchical equations of motion in real and imaginary time:
   Correlated initial states and thermodynamic quantities},}\ }\href {\doibase
  10.1063/1.4890441} {\bibfield  {journal} {\bibinfo  {journal} {J. Chem. Phys.}\ }\textbf {\bibinfo {volume} {141}},\ \bibinfo {pages}
  {044114} (\bibinfo {year} {2014})}\BibitemShut {NoStop}%
\bibitem [{\citenamefont {Koyanagi}\ \emph {et~al.}(2024)\citenamefont {Koyanagi},
   and\ \citenamefont {Tanimura}}]{Koyanagi2024jcp2}%
  \BibitemOpen
  \bibfield  {author} {\bibinfo {author} {\bibfnamefont {S.}~\bibnamefont
  {Koyanagi}} and\ \bibinfo {author}
  {\bibfnamefont {Y.}~\bibnamefont {Tanimura}},\ }\bibfield  {title}
  {\enquote {\bibinfo {title} {Thermodynamic quantum Fokker-Planck
  equations and their application to thermostatic Stirling engine},}\ }\href {\doibase
  10.1063/5.0225607} {\bibfield  {journal} {\bibinfo  {journal} {J. Chem. Phys.}\ }\textbf {\bibinfo {volume} {161}},\ \bibinfo {pages}
  {112501} (\bibinfo {year} {2024})}\BibitemShut {NoStop}%
\bibitem [{\citenamefont {Tanimura}(2015)\citenamefont {Tanimura}
   }]{Tanimura2015jcp}%
  \BibitemOpen
  \bibfield  {author} {\bibinfo {author}
  {\bibfnamefont {Y.}~\bibnamefont {Tanimura}},\ }\bibfield  {title}
  {\enquote {\bibinfo {title} {Real-time and imaginary-time quantum hierarchal FokkerPlanck equations},}\ }\href {\doibase
  10.1063/1.4916647} {\bibfield  {journal} {\bibinfo  {journal} {J. Chem. Phys.}\ }\textbf {\bibinfo {volume} {142}},\ \bibinfo {pages}
  {144110} (\bibinfo {year} {2015})}\BibitemShut {NoStop}%
\bibitem [{\citenamefont {Zhang}\ \emph {et~al.}(2008)\citenamefont {Zhang},
  \citenamefont {Hernandez}, and\ \citenamefont {Zhu}}]{Zhang2008ol}%
  \BibitemOpen
  \bibfield  {author} {\bibinfo {author} {\bibfnamefont {J.-P.}~\bibnamefont
  {Zhang}}, \bibinfo {author} {\bibfnamefont {G.}\ \bibnamefont {Hernandez}}, and\ \bibinfo {author}
  {\bibfnamefont {Y.-F.}~\bibnamefont {Zhu}},\ }\bibfield  {title}
  {\enquote {\bibinfo {title} {Slow light with cavity electromagnetically induced
  transparency},}\ }\href {\doibase
  10.1364/OL.33.000046} {\bibfield  {journal} {\bibinfo  {journal} {Opt. Lett.}\ }\textbf {\bibinfo {volume} {33}},\ \bibinfo {pages}
  {46} (\bibinfo {year} {2008})}\BibitemShut {NoStop}%
\bibitem [{\citenamefont {Seiler}\ \emph {et~al.}(2017)\citenamefont {Seiler},
  \citenamefont {Palato}, and\ \citenamefont {Kambhampati}}]{Seiler2017jcp}%
  \BibitemOpen
  \bibfield  {author} {\bibinfo {author} {\bibfnamefont {H.}~\bibnamefont
  {Seiler}}, \bibinfo {author} {\bibfnamefont {S.}\ \bibnamefont {Palato}}, and\ \bibinfo {author}
  {\bibfnamefont {P.}~\bibnamefont {Kambhampati}},\ }\bibfield  {title}
  {\enquote {\bibinfo {title} {Coherent multi-dimensional spectroscopy at optical frequencies in a single beam with optical readout},}\ }\href {\doibase
  10.1063/1.4990500} {\bibfield  {journal} {\bibinfo  {journal} {J. Chem. Phys.}\ }\textbf {\bibinfo {volume} {147}},\ \bibinfo {pages}
  {094203} (\bibinfo {year} {2017})}\BibitemShut {NoStop}%
\bibitem [{\citenamefont {Wang}\ \emph {et~al.}(2022)\citenamefont {Wang},
  \citenamefont {Qi}, \citenamefont {Pistore}, \citenamefont {Li}, \citenamefont {Agnew}, \citenamefont {Linfield}, \citenamefont {Davies}, \citenamefont {Tignon}, \citenamefont {Mangeney}, \citenamefont {Raki\ifmmode \acute{c}\else \'{c}\fi{}}, and\ \citenamefont {Dhillon}}]{Wang2022pra}%
  \BibitemOpen
  \bibfield  {author} {\bibinfo {author} {\bibfnamefont {F.-H.}~\bibnamefont
  {Wang}}, \bibinfo {author} {\bibfnamefont {X.-Q.}\ \bibnamefont {Qi}}, \bibinfo {author} {\bibfnamefont {V.}\ \bibnamefont {Pistore}}, \bibinfo {author} {\bibfnamefont {L.-H.}\ \bibnamefont {Li}}, \bibinfo {author} {\bibfnamefont {G.}\ \bibnamefont {Agnew}}, \bibinfo {author} {\bibfnamefont {E.}\ \bibnamefont {Linfield}}, \bibinfo {author} {\bibfnamefont {G.}\ \bibnamefont {Davies}}, \bibinfo {author} {\bibfnamefont {J.}\ \bibnamefont {Tignon}}, \bibinfo {author} {\bibfnamefont {J.}\ \bibnamefont {Mangeney}}, \bibinfo {author} {\bibfnamefont {A. D.}\ \bibnamefont {Raki\ifmmode \acute{c}\else \'{c}\fi{}}}, and\ \bibinfo {author}
  {\bibfnamefont {S. S.}~\bibnamefont {Dhillon}},\ }\bibfield  {title}
  {\enquote {\bibinfo {title} {Ultrafast Buildup Dynamics of Terahertz Pulse Generation in Mode-Locked Quantum Cascade Lasers},}\ }\href {\doibase
  10.1103/PhysRevApplied.18.064054} {\bibfield  {journal} {\bibinfo  {journal} {Phys. Rev. Appl.}\ }\textbf {\bibinfo {volume} {18}},\ \bibinfo {pages}
  {064054} (\bibinfo {year} {2022})}\BibitemShut {NoStop}%
\bibitem [{\citenamefont {Yan}\ \emph {et~al.}(2022)\citenamefont {Yan},
  \citenamefont {Revesz}, \citenamefont {Liang}, and\ \citenamefont {Li}}]{Yan2022pra}%
  \BibitemOpen
  \bibfield  {author} {\bibinfo {author} {\bibfnamefont {J.-L.}~\bibnamefont
  {Yan}}, \bibinfo {author} {\bibfnamefont {S.}\ \bibnamefont {Revesz}}, \bibinfo {author} {\bibfnamefont {D.-F.}\ \bibnamefont {Liang}}, and\ \bibinfo {author}
  {\bibfnamefont {H.-B.}~\bibnamefont {Li}},\ }\bibfield  {title}
  {\enquote {\bibinfo {title} {Broadband optical two-dimensional coherent spectroscopy of a rubidium atomic vapor},}\ }\href {\doibase
  10.1103/PhysRevA.105.052810} {\bibfield  {journal} {\bibinfo  {journal} {Phys. Rev. A}\ }\textbf {\bibinfo {volume} {105}},\ \bibinfo {pages}
  {052810} (\bibinfo {year} {2022})}\BibitemShut {NoStop}%
\bibitem [{\citenamefont {Geng}\ \emph {et~al.}(2014)\citenamefont {Geng},
  \citenamefont {Campbell}, \citenamefont {Bernu}, \citenamefont {Higginbottom}, \citenamefont {Sparkes}, \citenamefont {Assad},  \citenamefont {Zhang},  \citenamefont {Robins},  \citenamefont {Lam}, and\ \citenamefont {Buchler}}]{Geng2014njp}%
  \BibitemOpen
  \bibfield  {author} {\bibinfo {author} {\bibfnamefont {J.}~\bibnamefont
  {Geng}}, \bibinfo {author} {\bibfnamefont {G.~ T.}\ \bibnamefont {Campbell}}, \bibinfo {author} {\bibfnamefont {J.}\ \bibnamefont {Bernu}}, \bibinfo {author} {\bibfnamefont {D. B.}\ \bibnamefont {Higginbottom}}, \bibinfo {author} {\bibfnamefont {B.~M.}\ \bibnamefont {Sparkes}}, \bibinfo {author} {\bibfnamefont {S.~M.}\ \bibnamefont {Assad}},  \bibinfo {author} {\bibfnamefont {W.-P.}\ \bibnamefont {Zhang}},  \bibinfo {author} {\bibfnamefont {N.~P.}\ \bibnamefont {Robins}},  \bibinfo {author} {\bibfnamefont {P.~K.}\ \bibnamefont {Lam}} and\ \bibinfo {author}
  {\bibfnamefont {B.~C.}~\bibnamefont {Buchler}},\ }\bibfield  {title}
  {\enquote {\bibinfo {title} {Electromagnetically induced transparency and four-wave mixing in a cold atomic ensemble with large optical depth},}\ }\href {\doibase
  10.1088/1367-2630/16/11/113053} {\bibfield  {journal} {\bibinfo  {journal} {New J. Phys.}\ }\textbf {\bibinfo {volume} {16}},\ \bibinfo {pages}
  {113053} (\bibinfo {year} {2014})}\BibitemShut {NoStop}%
\end{thebibliography}
\providecommand{\noopsort}[1]{}\providecommand{\singleletter}[1]{#1}%

\end{document}